%% file: apssamp.tex
\pdfoutput=1
\documentclass[aps,prd,amsmath,floats,floatfix, twocolumn,
superscriptaddress,nofootinbib,showpacs,longbibliography]{revtex4-1}
\input{preamble}
\usepackage{physics}
\usepackage{aas_macros}
\usepackage{orcidlink}
\usepackage{enumitem}
\usepackage{subcaption}
\DeclareMathAlphabet{\mathpzc}{OT1}{pzc}{m}{it}

\begin{document}

\preprint{APS/123-QED}

\newcommand{\CITA}{\affiliation{Canadian Institute for Theoretical Astrophysics, University of Toronto, 60 St. George Street, Toronto, ON M5S 3H8, Canada}}
\newcommand{\IUCAA}{\affiliation{Inter University Centre for Astronomy and Astrophysics, Post Bag 4, Ganeshkhind, Pune - 411007, India}}
\newcommand{\Northwestern}{\affiliation{Center for Interdisciplinary Exploration $\&$ Research in Astrophysics (CIERA), Northwestern University, Evanston, IL 60208, USA}}
\newcommand{\TIFR}{\affiliation{Tata Institute of Fundamental Research, Homi Bhabha Road, Navy Nagar, Colaba, Mumbai 400005, India}}

\title{Profiling stellar environments of gravitational wave sources}

\author{Avinash Tiwari}
\IUCAA{}
\author{Aditya Vijaykumar}
\CITA{}
\author{Shasvath J. Kapadia}
\IUCAA{}
\author{Sourav Chatterjee}
\TIFR{}
\author{Giacomo Fragione}
\Northwestern{}

\date{\today}

\begin{abstract}

Gravitational waves (GWs) have enabled direct detections of compact binary coalescences (CBCs). However, their poor sky localisation and the typical lack of observable electromagnetic (EM) counterparts make it difficult to confidently identify their hosts, and study the environments that nurture their evolution. In this work, we show that {\it detailed} information of the host environment (e.g. the mass and steepness of the host potential) can be directly inferred by measuring the kinematic parameters (acceleration and its time-derivatives) of the binary's center of mass using GWs alone, without requiring an EM counterpart. 
We consider CBCs in various realistic environments such as globular clusters, nuclear star clusters, and active galactic nuclei disks to demonstrate how orbit and environment parameters can be extracted for CBCs detectable by ground- and space-based observatories, including the LIGO detector at A+ sensitivity, Einstein Telescope of the XG network, LISA, and DECIGO, {\it on a single-event basis}. 
These constraints on host stellar environments promise to shed light on our understanding of how CBCs form, evolve, and merge. 
\end{abstract}

\maketitle

\section{Introduction}\label{sec: Introduction}
A number of efforts trying to probe the provenance of compact binary coalescences (CBCs) detected with the LIGO-Virgo-KAGRA (LVK) gravitational wave (GW) detector network~\cite{LIGODetector, VirgoDetector, KAGRADetector, KAGRA:2021vkt} are currently underway (see, e.g., \cite{Mapelli:2021taw} for a review). 
Indeed, determining the environment in which the CBC formed, evolved, and merged is highly non-trivial for two primary reasons.
First, the GW waveform pertaining to the CBC does not always contain signatures of the CBC's formation channel. 
While statistical arguments based on estimates of the CBC's intrinsic source parameters could provide hints of its host environment, these cannot in general conclusively rule out all other possible environments in favor of the statistically preferred one \cite{Zevin:2020gbd, Pierra:2024fbl}. Second, the lack of observed EM counterparts for all but one of the detected CBC events ~\cite{PhysRevLett.119.161101, Abbott_2017:GW170817, KAGRA:2021vkt}, in tandem with poor localisation sky-areas spanning $\mathcal{O}(10)$ sq. deg. or larger, make it difficult to identify the host via direct observations \cite{Chen:2016tys}. The inclusion of additional detectors in current and forthcoming observing runs, such as KAGRA and LIGO-India \cite{Saleem:2021iwi}, as well as planned and proposed next-generation ground and space-based missions, will have markedly improved localisation sky-areas, even going below $\mathcal{O}(1)$ sq. deg. for space-based missions \cite{Zhang:2020hyx}. However, confident identification of the CBC's host will require arc- or sub-arc-second localisations, which will in general not be achievable even by future detectors, especially at large redshifts. 

Kinematic parameters (such as acceleration) associated with the motion of the center of mass (CoM) of the CBC, could serve as messengers of the CBC's environment. Past work has shown that a finite line-of-sight acceleration of the CBC's CoM will modulate the GW waveform with respect to one produced by a non-accelerated CBC ~\cite{Vijaykumar_2023, Bonvin2017, Yunes2011}. At leading order, these modulations occur at $-4$ post Newtonian (PN) order. Higher time derivatives of this acceleration (e.g. jerk, snap, etc) cause modulations that start at even lower PN orders (e.g. $-8$ PN, $-12$ PN, etc) \cite{Yunes2011}. Thus, detectors with sensitivities at low frequencies, such as space-based ones, are best suited to extract these kinematic parameters, although sufficiently large accelerations might be observable even with ground-based detectors \cite{Vijaykumar_2023}.  

Most proposed formation channels of stellar-mass CBCs suggest that these either underwent isolated evolution in the galactic field, or were assembled dynamically in dense stellar environments such as globular clusters (GCs), nuclear star clusters (NSCs), and active galactic nuclei (AGNs). These environments have characteristic properties, in particular, mass and potential profiles, which govern the kinematics of the CoMs of the CBCs they host, thereby imprinting themselves on the emanated GWs. In this paper, we demonstrate how these properties can be directly extracted from the CBC's GWs, thus effectively identifying the CBC's environment on a {\it single-event basis}. 

We calculate the corrections to the GW phase as a function of the kinematic parameters. We then evaluate the relationship between the kinematic parameters and the properties of the environment (e.g. steepness of the potential), as well as the orbit of the CoM of the binary. We are thus able to extract and constrain the corresponding parameters and explore the precision of their inference in various single-detector configurations, viz. a LIGO detector at A+ sensitivity \cite{aplus}, Einstein Telescope (ET) \cite{punturo2010} of the XG network \cite{Reitze:2019iox, punturo2010}, the LISA \cite{LISA:2017pwj} and DECIGO \cite{Sato:2017dkf} space-based detectors.

\section{Method}\label{sec: Method}
\subsection{Phase Correction}
The total mass of a binary $M = m_1 + m_2$ in the detector frame as inferred from the GW signal is related to the source frame total mass $M_{\mathrm{src}}$  as:
\begin{equation}
M = M_{\mathrm{src}}(1 + z_{\mathrm{cos}})( 1 + z_{\mathrm{dop}}) \qq{,}
\end{equation} 
where $z_{\mathrm{cos}}$ is the cosmological redshift, and $z_{\mathrm{dop}}$ is the Doppler shift caused by any constant line of sight (LOS) velocity (LOSV) of the CoM of the binary.
A constant LOSV is fully degenerate with the source frame mass. However, a time-varying LOSV results in a time-varying Doppler shift and induces a time-varying detector frame mass, allowing time derivatives of the LOSV to be measured. Modulations to the GW waveform appear at different PN orders depending on the order of the time derivative of LOSV causing the modulation. 

For a CoM with a finite LOS acceleration (LOSA), and higher LOSV ($v_l$) time derivatives, the detector frame mass becomes:
\begin{equation}
    \label{eq: Mz}
    M_z = M \left(1 + \sum_{n = 1}^{\infty} \Gamma_n (t_{\rm o} - t_{\rm c})^n\right)
\end{equation}
where $\Gamma_n \equiv \left. \frac{1}{c}\frac{1}{n!}\frac{d^n v_l}{dt_{\rm o} ^n} \right\vert_{t_{\rm o} = t_{\rm c}}$, $t_{\rm o} $ is the observation time, $t_{\rm c} $ is the time at coalescence, and $c$ is the speed of light. Eq.~\eqref{eq: Mz} is valid under the assumption that $| \Gamma_n (t_{\rm o} - t_{\rm c})^n |$ are $\ll$ 1. As a consistency check of this condition, we ensure that $|\Gamma_n (t_{\rm o} - t_{\rm c})^n | \leq 0.05$, where special relativistic effects may be neglected. 

The corresponding modulated inspiral GW waveform, in the stationary phase approximation, may be written as \cite{Buonanno:2009zt}:
\begin{equation}
    \label{eq: WF}
    \Tilde{h}(f) = \mathcal{A} f^{-7/6} \exp{i\left[ \Psi (f) + \sum_{n=1}^{\infty}\Delta \Psi_{-4n} (f;\Gamma_n) \right]}
\end{equation}
where $\mathcal{A} \propto \sqrt{\frac{5}{24}} \frac{\mathcal{M}^{5/6}}{D_{\rm L}}$ is the amplitude, $\mathcal{M}$ is the chirp mass, $D_{\rm L}$ is the luminosity distance, $\Psi (f)$ is the unperturbed phase given by Eq.~3.18 of \cite{Buonanno:2009zt}, and $\Delta \Psi_{-4n} (f; \Gamma_n)$ is a $-4n$ PN phase correction due to $n^{th}$ derivative of $v_l$ and at the leading order, it is given by:

\begin{equation}
    \label{eq: pcnth}
    \Delta \Psi_{-4n} (f) = -\frac{ 2^{-8 n-7} 5^{n+1} v^{-8 n-5} \left(-\frac{M}{\eta }\right)^n}{\eta  (n+1)} \Gamma_n
\end{equation}
where $v = (\pi M f)^{1/3}$, $\eta \equiv m_1m_2/(m_1 + m_2)^2$ is the symmetric mass ratio, and $f$ is the observed GW frequency. We consider the CBCs to be quasi-circular and non-spinning. Although CBCs at low frequencies are expected to have a residual eccentricity, the corresponding leading order LOSA phase corrections \cite{2024arXiv240305625S} are identical to the quasi-circular case.

Positive powers of a kinematic parameter can contribute to PN orders lower than the corresponding leading order. For example, the $-8$ PN phase correction $\Delta \Psi_{-8} (f)$ will have contributions from the LOS jerk (LOSJ) $\Gamma_2$ of the CoM, as well as the LOSA term proportional to $\Gamma_1^2$. We account for these corrections as well, down to $-24$ PN (see Appendix~\ref{subsec: phase_cor_5}), below which we do not expect PN terms to be well constrained even with detectors sensitive down to mili-hertz frequencies (see Figure~\ref{fig: pc_ac_BW} of the Appendix~\ref{sec: supp_fig}). 

For completeness, in Appendix~\ref{sec: amp_cor}, we also compute the amplitude corrections. However, we do not use them while calculating the Fisher Matrix. We do not expect this choice to impact our results, and we show this in Appendix~\ref{sec: supp_fig}  by considering one of the specific cases of this analysis. 
In particular, we show that the leading order (relative) contributions to the amplitude due to the derivatives of LOSV are small i.e., $\ll 1$ (see the right panel of Figure \ref{fig: pc_ac_BW} in Appendix~\ref{sec: supp_fig}).

\begin{figure}
    \centering
    \includegraphics[width = 0.95\linewidth]{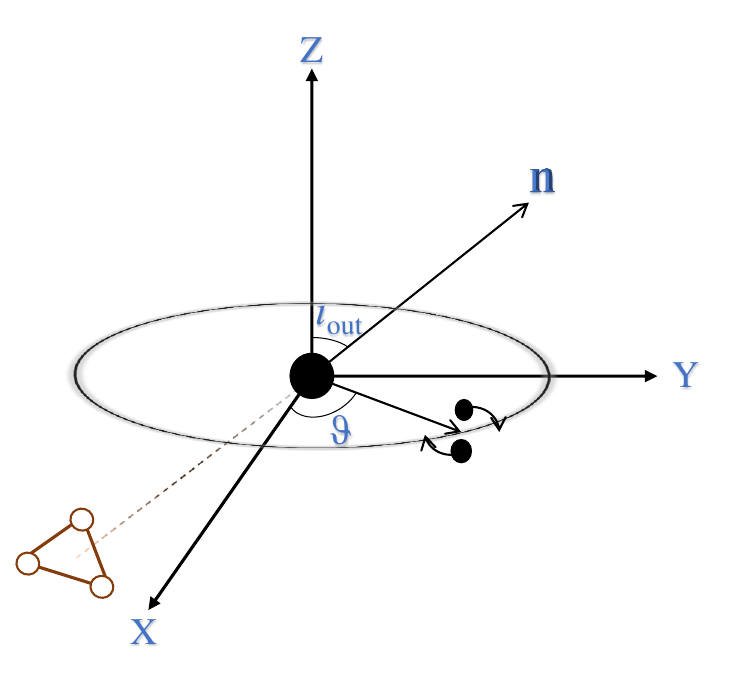}
    \caption{A schematic representation of a binary orbiting around an SMBH (or, any center of potential) in a circular orbit. Here, $\iota_{\mathrm{out}}$ is the angle between our LOS $\boldsymbol{n}$ and the angular momentum vector of the outer orbit, which is along $\boldsymbol{e}_z$. }
    \label{fig: co_ord_sys}
\end{figure}

\subsection{Kinematics of the CoM of the CBC}
To analyse the kinematics of the CoM of the CBC in these environments, we set up a Cartesian coordinate system in Figure \ref{fig: co_ord_sys} placed at the center of the gravitating source. Let $\iota_{\mathrm{out}}$ be the angle between our LOS and the angular momentum vector of the outer orbit, and let the projection vector of our LOS onto the outer orbital plane be along $\boldsymbol{e}_y$. Then:
\begin{equation}
    \label{eq: unitvect}
    \begin{aligned}
        \boldsymbol{n} &= \sin\iota_{\mathrm{out}} \boldsymbol{e}_y + \cos\iota_{\mathrm{out}} \boldsymbol{e}_z \\
        \boldsymbol{e}_r (\vartheta) &= \cos\vartheta \boldsymbol{e}_x + \sin\vartheta \boldsymbol{e}_y \\
        \boldsymbol{e}_{\vartheta} (\vartheta) &= - \sin\vartheta \boldsymbol{e}_x + \cos\vartheta \boldsymbol{e}_y
    \end{aligned}
\end{equation}
where $\vartheta$ is the angular location of the binary in the orbital plane with respect to the x-axis, and $r$ is the radial location.

Let us further define a vector $\boldsymbol{\mathcal{V}} (r, \vartheta)$ such that:
\begin{equation}
    \label{eq: vect}
    \boldsymbol{\mathcal{V}} (r, \vartheta) \equiv \mathcal{V}_r (r) \boldsymbol{e}_r (\vartheta) + \mathcal{V}_{\vartheta} (r) \boldsymbol{e}_{\vartheta} (\vartheta)
\end{equation}
where $r \equiv r(t)$, $\vartheta \equiv \vartheta(t)$, and $\boldsymbol{e}_r (\vartheta)$ and $ \boldsymbol{e}_{\vartheta} (\vartheta)$ are the unit vectors. Then observing that $\Dot{\boldsymbol{e}_r} = \Dot{\vartheta}\boldsymbol{e}_{\vartheta}$ and $\Dot{\boldsymbol{e}_{\vartheta}} = - \Dot{\vartheta}\boldsymbol{e}_r$, we can write:
\begin{equation}
    \label{eq: vectder}
    \Dot{\boldsymbol{\mathcal{V}}} (r, \vartheta) = (\mathcal{V}'_r \Dot{r} - \mathcal{V}_{\vartheta} \Dot{\vartheta}) \boldsymbol{e}_r  + (\mathcal{V}'_{\vartheta} \Dot{r} + \mathcal{V}_r \Dot{\vartheta}) \boldsymbol{e}_{\vartheta} 
\end{equation}
where $\cdot \equiv \frac{d}{dt}$ and $' \equiv \frac{d}{dr}$. Using this relation iteratively, we calculate different kinematic parameters of the CoM of the binary and then project them onto our LOS ($\boldsymbol{n}$), which can be written as:
\begin{equation}
    \label{eq: vectder_proj}
    \Dot{\mathcal{V}}_l \equiv \Dot{\boldsymbol{\mathcal{V}}} \cdot \boldsymbol{n} = [(\mathcal{V}'_r \Dot{r} - \mathcal{V}_{\vartheta} \Dot{\vartheta}) \sin \vartheta +  (\mathcal{V}'_{\vartheta} \Dot{r} + \mathcal{V}_r \Dot{\vartheta}) \cos \vartheta] \sin \iota_{\mathrm{out}}.
\end{equation}
We provide the derivation as well as the mathematical expressions of the LOS components in the Appendix~\ref{sec: kin_param_derivation}.

The $\Gamma_n$'s will, in general, be functions of environment and outer-orbit parameters, such as the steepness of the gravitational potential profile, and the location of the CBC with respect to the center of the potential, at coalescence. For instance, a circular orbit of radius $R$ in a Keplerian potential around an SMBH of mass $M_{\rm SMBH}$ would have $\Gamma_n \sim \omega^{n+1} R$, where $\omega^2 \sim M_{\rm SMBH} / R^3$ (see Appendix~\ref{subsec: app_CO}). 

\subsection{Fisher Matrix Analysis}\label{subsec: fma}
To assess our ability to constrain the environment and
outer-orbit parameters in various observing scenarios, we adopt a Fisher Matrix approach~\cite{CutlerFlanagan}. We construct the Fisher matrix in $\{\ln D_{\rm L}, \ln \mathcal{M}, \ln \eta, \Gamma_1, \Gamma_2\}$. We add or remove $\Gamma_3$, $\Gamma_4$, $\Gamma_5$, and $\Gamma_6$ to this set, as required, depending on the number of parameters that characterize the environment and outer-orbit. The PN corrections to the GW phase due to the kinematic parameters are rewritten in terms of the environment and outer-orbit parameters. The Fisher matrix is then inverted to yield a Covariance Matrix, which provides the Gaussian approximation to the GW likelihood. The environment parameters are sampled from this likelihood using \texttt{emcee}~\cite{emcee}, and the relative r.m.s errors on these parameters are estimated. 

While calculating the covariance matrix, we re-write Eq. (2.7) of \cite{CutlerFlanagan} as:
\begin{equation*}
    \Gamma_{ij} = 4 \mathrm{Re} \int_{f_l}^{f_h}  \left(\frac{\partial \ln \Tilde{h}(f)}{\partial \theta_i}\right)^\ast   \frac{\partial \ln \Tilde{h}(f)}{\partial \theta_j}  |\Tilde{h}(f)|^2 \frac{df}{S_n(f)}
\end{equation*}
where $f_l$ and $f_h$ are the frequencies when the signal enters the detector band and when it leaves the band, respectively. To avoid dealing with large numbers and numerical over/under-flows, let us define $\nu \equiv f/f_0$ and $S(\nu f_0) \equiv S_n(\nu f_0)/S_0$. Then using Eq. \eqref{eq: WF}, we can write the above Eq. as $\Gamma_{ij} = 4 |\mathcal{A}|^2 \frac{f_0^{-4/3}}{S_0} I_{ij}$ where:
\begin{equation}
    \label{eq: FM_red}
    I_{ij} = \mathrm{Re} \int_{\nu_l}^{\nu_h}  \left(\frac{\partial \ln \Tilde{h}(\nu f_0)}{\partial \theta_i}\right)^\ast   \frac{\partial \ln \Tilde{h}(\nu f_0)}{\partial \theta_j} \frac{\nu^{-7/3}}{S(\nu f_0)} d\nu 
\end{equation}
$\nu_l = f_l/f_0$, and $\nu_h = f_h/f_0$. We can then write the covariance matrix as:
\begin{equation}
    \label{eq: CM}
    C = \boldsymbol{\Gamma}^{-1} = \frac{S_0 f_0^{4/3}}{4 |\mathcal{A}|^2} \boldsymbol{I}^{-1}
\end{equation}

The quantities $f_0$ and $S_0$, in general, can be chosen to be any number that makes the inversion of the matrix $\boldsymbol{I}$ precise. However, here we choose them to be (roughly) the frequency at which the detector is most sensitive, and the corresponding value of the PSD, respectively. We further assume $4$ years of observation time for DECIGO and LISA, and the full sensitivity band of A+ and ET. Following \cite{PhysRevD.71.084025}, we then set a sensitivity band of $[10^{-2}, 10]\, {\rm Hz}$ ($[10^{-4}, 1]\, {\rm Hz}$) for DECIGO \cite{PhysRevD.83.044011, PhysRevD.95.109901} (LISA \cite{Robson_2019}), while for A+ and ET, we use the PSDs provided in~\cite{A_plus_psd} and ~\cite{ET_psd, Hild:2010id}, respectively. 

The kinematic parameters in (\ref{sec: phase_cor}) are in the observer's frame while the relations derived in the subsections (\ref{subsec: app_CO} - \ref{subsec: app_PP}) are in the source (unperturbed) frame. For a cosmological redshift $z_{\rm cos}$, the source frame time and observer frame time are related as $t_u = t_{\rm o} / (1 + z_{\rm cos})$. This leads to the rescaling of the kinematic parameters $\Gamma_n$ by a factor\footnote{It can be seen straightforwardly that \[\Gamma_{n,s} t_s^n = \Gamma_{n,s} \frac{t_{\rm o} ^n}{(1 + z_{\rm cos})^n} = \Gamma_{n,o} t_{\rm o} ^n \] where $\Gamma_{n,o} \equiv \Gamma_{n,u}/(1 + z_{\rm cos})^n$} $1/(1 + z_{\rm cos})^n$. Therefore, strictly, estimating errors on the source-frame kinematic parameters will need to incorporate errors on the estimation of $z_{\rm cos}$, which can be acquired from the Fisher matrix analysis posterior on $D_{\rm L}$ and assuming standard cosmology. However, since we consider CBCs at low redshifts ($\sim 0.022$ and $\sim 0.198$ corresponding to $100\, {\rm Mpc}$ and $1\, {\rm Gpc}$, respectively), we neglect errors on $z_{\rm cos}$, i.e., assume it to be perfectly measured, and use the rescaled $\Gamma_n$'s in calculations as well as during the sampling with true value of $z_{\rm cos}$.

\subsection{Signal-to-Noise Ratio (SNR)}\label{sec: SNR_cal}
The optimal SNR is given by:
\begin{equation}
\varrho = \sqrt{4 |\mathcal{A}|^2 \mathcal \int_{f_l}^{f_h} \frac{f^{-7/3}}{S_n(f)} df}
\end{equation}
where $f_l$ and $f_h$ are the detector bandwidth's lower and upper frequencies. $S_n(f)$ is the noise power spectral density (PSD). We assume face-on inner orbits which maximizes the SNR. When the binary is inclined, the SNRs will be reduced by a factor $\frac{\mathcal{Q} (\iota)}{\mathcal{Q} (\iota = 0)}$ where:
\begin{equation}
\mathcal{Q}(\iota) = \sqrt{\left( \frac{1 + \cos^2\iota}{2}\right)^2 + \cos^2\iota}   
\end{equation}
and $\iota$ is the inclination of the (inner) binary relative to the LOS \cite{Robson_2019}.


\begin{figure*}[ht!]
    \centering
    \includegraphics[width = 0.9\linewidth]{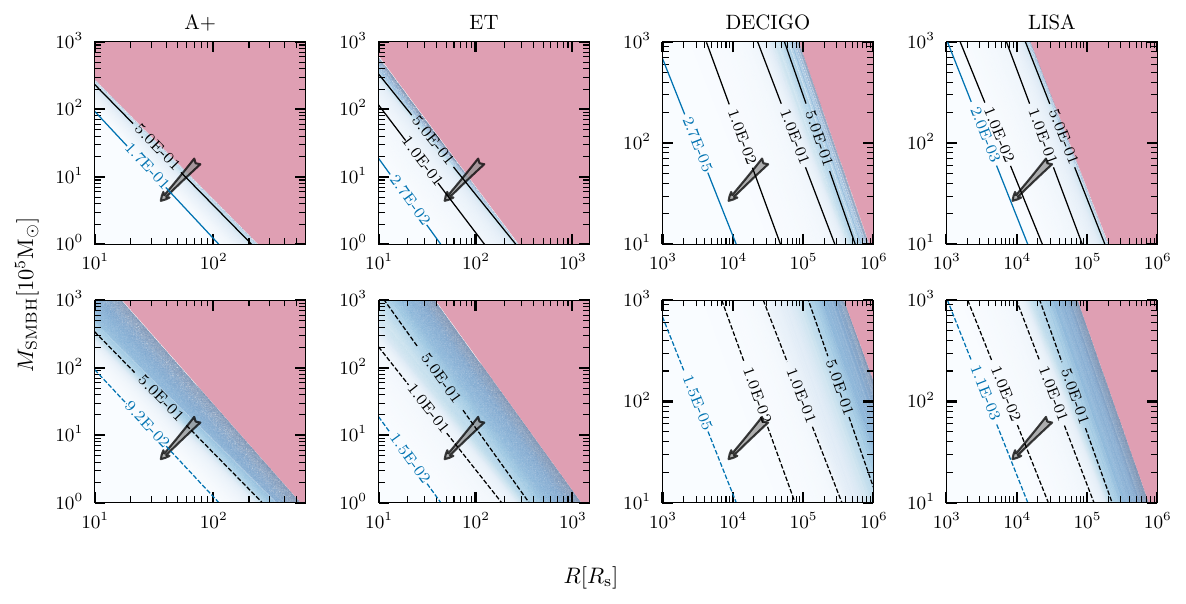}
    \caption{{\bf Circular outer-orbit around an SMBH:} The precision of the extraction of parameters over a grid of SMBH masses $M_{\rm SMBH}$ and radii of outer orbit $R$ for a binary system of component masses (\textit{from left to right}) $5-1.4 \,{\rm M}_{\odot}$ at $\mathrm{100Mpc}$ in A+, $30-30 \,{\rm M}_{\odot}$ at $100\mathrm{Mpc}$ in ET, and $100-100 \,{\rm M}_{\odot}$ at $1\mathrm{Gpc}$ in DECIGO and LISA, with $\cos \vartheta = 0.9$. \textit{Top Panels}: Relative error in the extraction of mass of SMBHs. The `solid' lines are the lines of constant $\frac{\Delta M_{\rm SMBH}}{M_{\rm SMBH}}$. \textit{Bottom Panel}: Relative error in the extraction of radii of the outer orbit. The `dashed' lines are the lines of constant $\frac{\Delta R}{R}$. The pink regions correspond to the parameter space where either the relative errors are greater than 1 or we do not sample them due to the non-measurability of the kinematic parameters. The blue lines demarcate the regions (lower left) where we do not perform the Fisher matrix analysis, for, in these regions, the assumption $| \Gamma_n (t_{\rm o} - t_{\rm c})^n | \ll 1$ is no longer valid. The arrows point towards the region of increasing precision in the inference of parameters.}
    
    \label{fig: CO_SMBH}
\end{figure*}

\section{Results}\label{sec: results}
We present results for three CBC environments: vicinity of supermassive black hole (SMBH, Keplerian potential), globular cluster (GC, Plummer \cite{Plummer1911} profile), nuclear star cluster, (NSC, Bahcall-Wolf \cite{Hoang_2018} profile); three orbital shapes: circular, elliptic and effective Keplerian --- a bound orbit that is not necessarily closed; and four single-detector configurations: LIGO at A+ sensitivity, ET, DECIGO, and LISA. We do not consider outer orbits around stellar mass BHs, which, for numerous configurations, could be chaotic. Such systems are analyzed in \cite{2024arXiv240305625S, 2024arXiv240804603H} by focusing on the phase-shifts due to the R\o mer delay. While circular orbits are possible in all attractive potentials, closed elliptical orbits are only possible in attractive potentials with $1/r$ and $1/r^2$ profiles, where $r$ is the radial coordinate in the orbital plane. Moreover, only those orbits with a changing $r$ will be sensitive to the steepness of the potential profile. Thus, for Bahcall-Wolf (BW) and Plummer potentials, we only consider effective Keplerian orbits by evolving both $r$ and $\vartheta$ (angular coordinate) over the duration of the GW signal within the frequency band of the detectors. Additionally, we do not consider the CBC in GC and NSC environments for LIGO at A+ sensitivity and ET. This is because the signal duration will be very small, and for GC case kinematic parameters will be too small to be measurable; while for NSC case even though kinematic parameters can be measurable if the CBC is very close to the centre, the potential of SMBH will be dominating and hence the scenario will be similar to the CBC around an SMBH case.

\begin{figure*}[ht!]
    \centering
    \includegraphics[width = 0.9\linewidth]{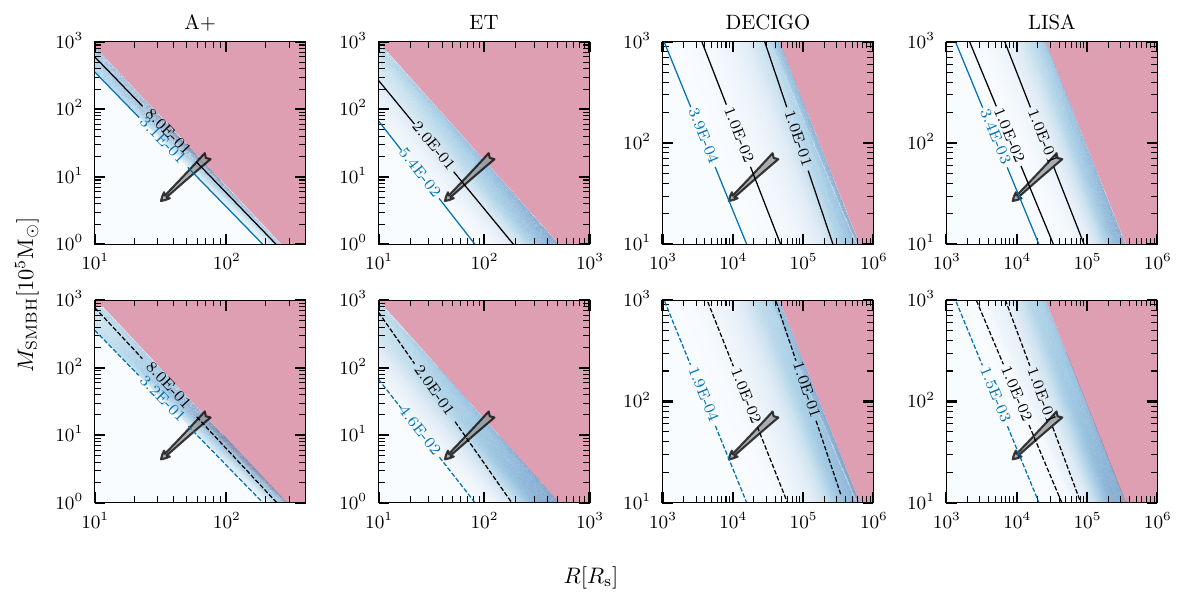}
    \caption{{\bf Eccentric outer-orbit around an SMBH:} The precision of the extraction of parameters over a grid of SMBH masses $M_{\rm SMBH}$ and semi-major axes $R$ of the outer orbit for a binary system of component masses (\textit{from left to right}) considered in Figure \ref{fig: CO_SMBH} with $e = 0.5$ and $x \approx 1.25$. $x$ was drawn randomly from a uniform distribution between $\frac{1}{1-e}$ and $\frac{1}{1+e}$. \textit{Top Panel}: Relative error in the extraction of mass of SMBHs. The `solid' lines are the lines of constant $\frac{\Delta M_{\rm SMBH}}{M_{\rm SMBH}}$. \textit{Bottom Panel}: Relative error in the extraction of the location of binary in the outer orbit. The `dotted' lines are the lines of constant $\frac{\Delta x}{x}$. The pink regions, the blue lines, and the arrows have the same meaning as in Figure \ref{fig: CO_SMBH}.}
    \label{fig: EO_SMBH}
\end{figure*}

\subsection{Circular Orbit around an SMBH}
For binaries in a \textit{circular orbit} around an SMBH with Schwarzschild radius $R_{\rm s}$, the kinematic parameters along our LOS depend on the mass of the SMBH ($M_{\rm SMBH}$), the radius of the circular orbit ($R$), and $\cos \vartheta$ (see Figure~\ref{fig: co_ord_sys}). We use the first three derivatives of LOSV, viz. LOSA, LOSJ, and LOS snap (LOSS), to uniquely determine these quantities. $M_{\rm SMBH}$ and $R$ will still be degenerate with $\iota_{\mathrm{out}}$ and this degeneracy can not be broken, even if we consider one more derivative. Similar degeneracies exist between environment/orbit parameters and $\iota_{\mathrm{out}}$ for all scenarios considered in this work. We set $\sin \iota_{\mathrm{out}} = 1$ throughout; thus, the relative errors in the inference of environment and orbit parameters should be thought of as lower limits. An alternative would have been to assume an uninformative prior on the inclination, which would increase the uncertainty in the inference of environment and orbit parameters by a factor of a few. 

Figure \ref{fig: CO_SMBH} shows the relative errors in the measurements of $M_{\rm SMBH}$ (top panel) and $R$ (bottom panel) over a $M_{\rm SMBH} - R$ grid. We denote the inference of a parameter as ``successful'' if the relative r.m.s. error is less than unity. For completeness, we also give the optimal signal-to-noise ratios (SNRs)  (see~\ref{sec: SNR_cal}) of the GW events we consider. We find that:
\begin{enumerate}
    \item At A+ sensitivity, we can successfully extract the mass of a $M_{\mathrm{SMBH}} = 10^5 \, {\rm M}_{\odot}$ SMBH and localise a $5-1.4 \, {\rm M}_{\odot}$ binary having optimal SNR $\sim 145$ corresponding to a luminosity distance $D_{\rm L} = 100 \,{\rm Mpc}$, out to $R \sim 200 \,R_{\rm s}$.
    \item With ET, we can successfully extract the mass of the same SMBH and localise a $30-30 \,{\rm M}_{\odot}$ binary having optimal SNR $\sim 8248$ corresponding to $D_{\rm L} = 100 \,{\rm Mpc}$, out to $R \sim 250 \,R_{\rm s}$, and within a precision of $\sim 10\%$ up to $R \sim 100 \,R_{\rm s}$.
    \item We can successfully infer the mass of an SMBH as large as $10^8 \,{\rm M}_{\odot}$, and localise a $100-100 \,{\rm M}_{\odot}$ binary having optimal SNR $\sim 10782 \; (14)$ corresponding to $D_{\rm L} = 1\, {\rm Gpc}$, out to $\sim 10^5 \,R_{\rm s}$ ($\sim 1.5 \times 10^4 \,R_{\rm s}$) with DECIGO (LISA).
\end{enumerate} 
Furthermore, for tight orbits around low mass black holes we can extract these quantities with sub-percent precision. All results in this analysis assume $\cos \vartheta$ to be 0.9. The angular location of the binary in the orbital plane at coalescence is a nuisance parameter that needs to be marginalized out when inferring other more astrophysically relevant parameters such as the radial location of the binary in the environment, and the environment's potential profile. Given the spherically symmetric nature of the potentials considered in this work, we do not expect the choice of this value to drastically change the inference of environment and orbit parameters. Nevertheless, to ascertain this, we show corner plots of the recovery of kinematic and environmental parameters in Figure~\ref{fig: CO_cos_theta_0.5} of the Appendix~\ref{sec: supp_fig}. 
We assume a $30-30 \, {\rm M}_{\odot}$ binary in circular orbit of radius $25 \, R_{\rm s}$ around a $10^6 \, {\rm M}_{\odot}$ SMBH, at $100\, {\rm Mpc}$, in ET, with $\cos \vartheta = 0.5$.  As with the $\cos \vartheta = 0.9$ case,  all of the parameters are recovered well within the posterior.

\begin{figure*}[ht!]
    \centering
    \includegraphics[width = 0.95\linewidth]{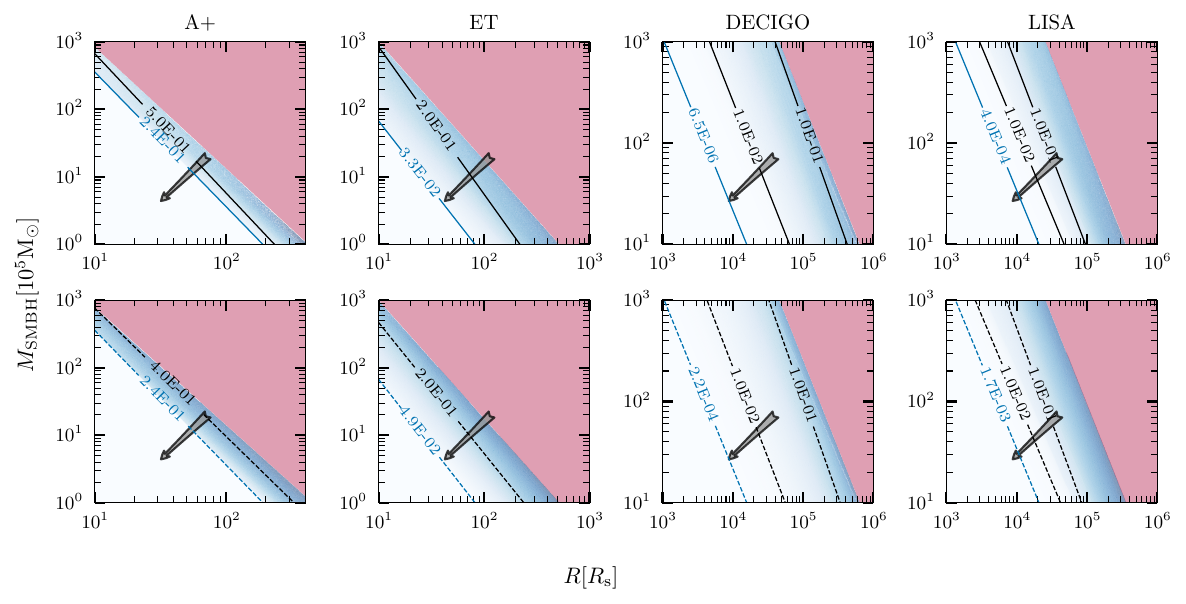}
    \caption{The relative errors in the semi-major axis and eccentricity of the outer orbit over a grid of SMBH masses $M_{\mathrm{SMBH}}$ and semi-major axis $R$ for the binary systems and detector configurations considered in Figure~\ref{fig: EO_SMBH} of the paper. \textit{Top Panels}: Relative error in the semi-major axis. The `solid' lines are the lines of constant $\frac{\Delta R}{R}$. \textit{Bottom Panels}: Relative error in the eccentricity. The `dashed' lines are the lines of constant $\frac{\Delta e}{e}$. The pink regions, the blue lines, and the arrows have the same meaning as in Figure \ref{fig: CO_SMBH}.}
    \label{fig: R_ecc_EO}
\end{figure*}

\subsection{Elliptical Orbit around an SMBH}
For binaries in an \textit{elliptical orbit} around an SMBH, the kinematic parameters along our LOS depend on $M_{\rm SMBH}$, semi-major axis ($R$), eccentricity ($e$), and $x \equiv R/r$ where $r$ is distance of the CoM of the binary from the gravitating center, at $t_{\rm c} $.

Therefore, together with LOSA, LOSJ, and LOSS, we will need one more derivative of LOSV: LOS crackle (LOSC), to uniquely determine these quantities (see Appendix~\ref{subsec: app_EO} for the expressions of kinematic parameters). 

Figure \ref{fig: EO_SMBH} shows the relative errors in the measurements of $M_{\rm SMBH}$ (top panel) and $x$ (bottom panel) over a $M_{\rm SMBH} - R$ grid. We find the following:

\begin{enumerate}
    \item With LIGO at A+ sensitivity, we can successfully extract the mass of a $10^5 \,{\rm M}_{\odot}$ SMBH and localise a $5-1.4 \,{\rm M}_{\odot}$ binary having optimal SNR $\sim 145$ corresponding to $D_{\rm L} = 100 \,{\rm Mpc}$, in an elliptical orbit of eccentricity $0.5$, up to a semi-major axis $R \sim 250 \,R_{\rm s}$.
    \item With ET, we can successfully extract the mass of the same SMBH and localise a $30-30 \,{\rm M}_{\odot}$ binary having optimal SNR $\sim 8248$ corresponding to $D_{\rm L} = 100 \,{\rm Mpc}$, in an elliptical orbit of eccentricity $0.5$, up to a semi-major axis $R \sim 500 \,R_{\rm s}$, and within a precision of $\sim 20\%$ up to $R \sim 200 \,R_{\rm s}$.
    \item We can successfully infer the masses of SMBHs as massive as $10^8 \,{\rm M}_{\odot}$, and localise a $100-100 \,{\rm M}_{\odot}$ binary having optimal SNRs $\sim 10782$ and $\sim 14$, corresponding to $D_{\rm L} = 1\, {\rm Gpc}$, in an elliptical orbit of eccentricity $0.5$, up to a semi-major axis $R \sim 5 \times 10^4 \,R_{\rm s}$ and $\sim 3 \times 10^4 \,R_{\rm s}$ with DECIGO and LISA, respectively.
\end{enumerate}

The chosen value of $x$ is a random draw from a uniform distribution between $\frac{1}{1-e}, \frac{1}{1+e}$. We also find that, for certain portions of the parameter space, we can achieve sub-percent precision in DECIGO (see lower left regions of the parameter space in Figure \ref{fig: EO_SMBH}). 

Figure \ref{fig: R_ecc_EO} shows the relative errors in the semi-major axis ($R$) and eccentricity ($e$) of the outer orbit over a grid of SMBH masses $M_{\mathrm{SMBH}}$ and $R$, with $e = 0.5$. The binary systems and detector configurations considered are the same as those in Figure~\ref{fig: EO_SMBH}. We can successfully extract $R$ and $e$, around a $10^5\, {\rm M}_{\odot}$ SMBH, out to $R \sim 400\, R_{\rm s}$ with A+, and $R \sim 500\, R_{\rm s}$ with ET. We can do the same around a $10^6\, {\rm M}_{\odot}$ SMBH out to $R \sim 7 \times 10^5\, R_{\rm s}$ with DECIGO, and $R \sim 10^6 \, R_{\rm s}$ with LISA. Furthermore, it is possible to extract these quantities within a sub-percent level of accuracy in certain portions of the parameter space (the lower left regions of the panels of Figure~\ref{fig: R_ecc_EO}). 

Note that we have verified that the CBCs considered in this paper will not be tidally disrupted by the SMBH, as the minimum semimajor axis required to avoid tidal disruption is much smaller than the values considered here.

\subsection{Effective Keplerian Orbit in BW potential}
For binaries in NSCs with a BW potential profile, the kinematic parameters projected along the LOS will depend on $M_{\rm MBH}$, $r$, steepness of the potential profile ($\alpha$), radial ($v_r$) and angular ($\Omega$) speeds of the CBC's CoM at coalescence, and $\cos \vartheta$. Hence, together with LOSA, LOSJ, LOSS, and LOSC, we will need two more derivatives of LOSV to break all degeneracies among the inferred parameters.

\begin{figure}
    \centering
    \includegraphics[width = 0.9\linewidth]{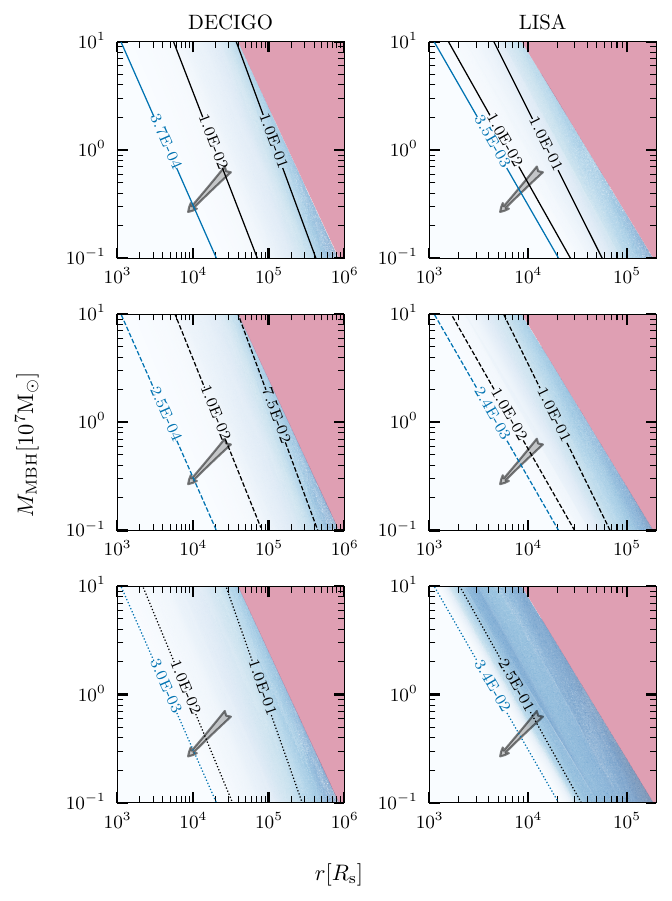}
    \caption{{\bf Effective Keplerian outer-orbit in an NSC with a BW profile:} The precision of the extraction of parameters over a grid of MBH masses $M_{\rm MBH}$ and merger distance $r$ for a binary system of component masses (\textit{from left to right}): $30-30\, {\rm M}_{\odot}$ and $100-100\, {\rm M}_{\odot}$ at $1{\rm Gpc}$ in DECIGO and LISA, respectively, for $\alpha = 1.75$ and $\cos \vartheta = 0.9$ with radial and tangential velocities: $v_r = 250\, {\rm km/s}$ and $r \Omega = 100\, {\rm km/s}$\footnote{These numbers are consistent with observed proper motions of stars in the Galactic NSC (see Figure 6 of \cite{2009A&A...502...91S}).}, respectively. \textit{Top Panel}: Relative error in the mass of the MBH. The `solid' lines are the lines of constant $\frac{\Delta M_{\rm MBH}}{M_{\rm MBH}}$.  \textit{Middle Panel}: Relative error in the distance of the binary from the NSC's center. The `dashed' lines are the lines of constant $\frac{\Delta r}{r}$. \textit{Bottom Panel}: Relative error in the steepness parameter $\alpha$. The `dotted' lines are the lines of constant $\frac{\Delta \alpha}{\alpha}$. The pink regions, the blue lines, and the arrows have the same meaning as in Figure \ref{fig: CO_SMBH}.}
    \label{fig: BW}
\end{figure}

Figure \ref{fig: BW} shows the relative errors in the measurements of $M_{\rm MBH}$ (top panel), $r$ (middle panel), and $\alpha$ (bottom panel) over a grid of MBH masses and the distances of the merger. We find that:
\begin{enumerate}

    \item With DECIGO, we can extract the mass of a $10^{6} \,{\rm M}_{\odot}$ MBH within a precision of $\sim 10\%$ together with $\alpha$, and localise a $30-30 \,{\rm M}_{\odot}$ binary having optimal SNR $\sim 3951$ corresponding to $D_{\rm L} = 1 \, {\rm Gpc}$, with a sub-percent precision out to $\sim 3 \times 10^5 \,R_{\rm s}$.
    \item With LISA, we can extract the mass of a $10^{6} \,{\rm M}_{\odot}$ MBH within a precision of $\sim 25\%$ together with $\alpha$ and localise a $100-100 \,{\rm M}_{\odot}$ binary having optimal SNR $\sim 14$ corresponding to $D_{\rm L} = 1 \, {\rm Gpc}$, with a sub-percent precision out to $\sim 3 \times 10^4 \,R_{\rm s}$.
\end{enumerate}
It is clear from Figure \ref{fig: BW} that we can extract $M_{\rm MBH}$ and $r$ with sub-percent precision for certain parts of this parameter space, for both DECIGO and LISA. However, we cannot do so for $\alpha$ with LISA, for the fiducial CBC we consider. This can be understood as follows. Larger magnitudes of kinematic parameters increase our ability to constrain environment parameters. Getting closer to the MBH does indeed increase these magnitudes; however, the MBH potential starts to dominate over the potential with exponent $\alpha$. As a result, the precision of $\alpha$ gets suppressed. There is therefore a sweet spot in the $M_{\rm MBH}-r$ parameter space where the precision on the inference of $\alpha$ is optimized. This spot changes, depending on intrinsic parameters of the CBC, including its orbital parameters. Figure \ref{fig: BW} does not clearly show this sweet spot because it lies within the highly-relativistic domain where our analysis is not carried out. 

As an illustration, in Figure~\ref{fig: EnvP_BW_DECIGO} of  Appendix~\ref{sec: supp_fig}, we have shown the recoveries of the mass of the central massive black hole $M_{\mathrm{MBH}}$, the radial location of the CBC in the outer orbit $r$, and the steepness of the gravitational potential $\alpha$, for a $30-30\, \mathrm{M}_{\odot}$ binary at $1\,\mathrm{Gpc}$, in DECIGO.

\subsection{Effective Keplerian Orbit in Plummer potential}
For binaries in GCs with a \textit{Plummer potential} \cite{Plummer1911, 1974A&A....37..183A} profile, kinematic parameters along our LOS will depend on the total mass of the cluster ($M_0$), $r$, Plummer radius ($a_{\mathrm{p}}$), $v_r$, $\Omega$, and $\cos \vartheta$. Therefore, similar to the BW case, we will need six derivatives of LOSV to uniquely determine parameters. However, the Plummer profile being shallower than the BW profile, recovering LOSC and higher time derivatives of LOSV is not possible for the range of parameters we consider. Hence, we only evaluate the LOSA, LOSJ, and LOSS (see \ref{subsec: app_PP} for the expressions) when constructing the GW phase modulations.

\begin{figure*}
    \centering
    \includegraphics[width = 0.95\linewidth]{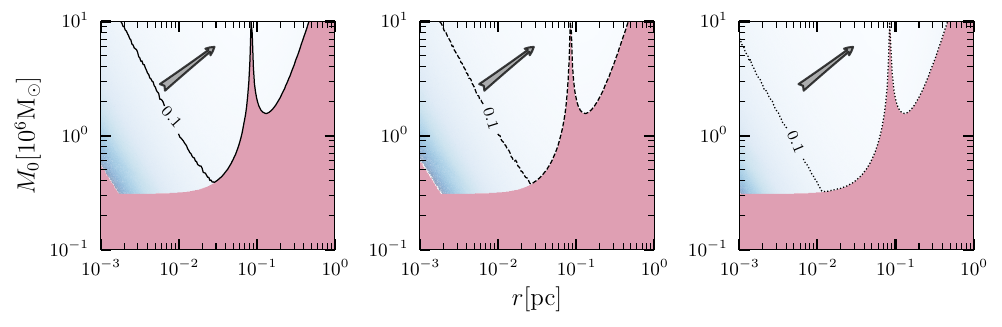}
    \caption{{\bf Effective Keplerian outer-orbit in a GC with Plummer profile:} The precision of the extraction of parameters over a grid of cluster masses $M_0$ and merger distance $r$ for a binary system of component masses $30-30 \,{\rm M}_{\odot}$ at $100\mathrm{Mpc}$ in DECIGO with $\cos \vartheta = 0.25$. \textit{Left Panel}: Relative error in the total mass of the cluster. The `solid' lines are the lines of constant $\frac{\Delta M_0}{M_0}$. \textit{Middle Panel}: Relative error in the distance of the binary from the GC's center. The `dashed' lines are the lines of constant $\frac{\Delta r}{r}$. \textit{Right Panel}: Relative error in the Plummer radius where $a_{\rm p} = 0.1\, {\rm pc}$. The `dotted' lines are the lines of constant $\frac{\Delta a_{\mathrm{p}}}{a_{\mathrm{p}}}$. The pink regions and the arrows have the same meaning as in Figure \ref{fig: CO_SMBH}.}
    \label{fig: PP}
\end{figure*}

Figure~\ref{fig: PP} shows the relative errors in the measurements of $M_{0}$ (top panel), $r$ (middle panel), and $a_{\mathrm{p}}$ (bottom panel) over a grid of $M_0$ and $r$ for $a_{\rm p} = 0.1 \,{\rm pc}$. We find that with DECIGO, we can successfully extract $M_0$, $a_{\rm p}$, and localise a $30-30\, \rm M_{\odot}$ binary having optimal SNR $\sim 39513$ corresponding to $D_{\rm L} = 100 \,{\rm Mpc}$, for total cluster masses greater than $\sim 3 \times 10^5 \,{\rm M}_{\odot}$.

Even though we can successfully constrain $M_0$, $r$, and $a_{\mathrm{p}}$ out to $r \sim 0.4\mathrm{pc}$, it still requires the cluster to be as heavy as $10^7 \,{\rm M}_{\odot}$ which is well above the most massive known GC~\cite{2013MNRAS.429.1887D}. 
With LISA, LOSJ is only measurable for a very limited portion of the parameter space we consider. Therefore, we do not show the corresponding results. 

\subsection{Stealth Bias}
\begin{figure}
    \centering
    \includegraphics[scale=0.535]{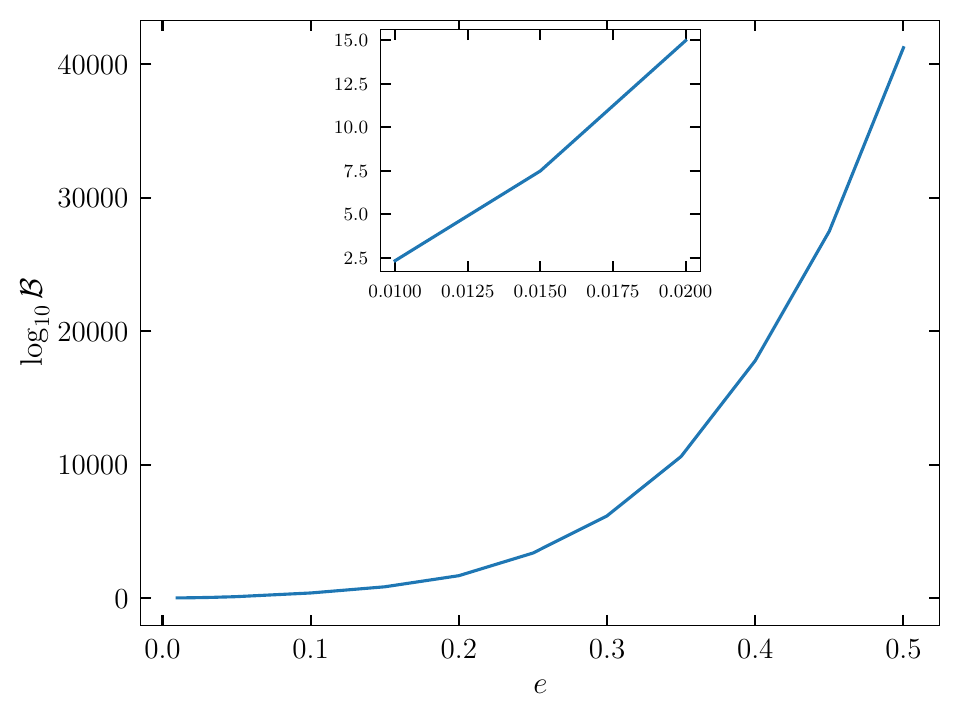}
    \caption{The variation of ($\log_{10}$) Bayes Factor $\mathcal{B}$ in favor of a non-zero eccentricity with the eccentricity of the outer orbit $e$ for the scenario considered in Figure \ref{fig: stealth_bias}. The inset plot shows the variation of the same at low eccentricities of the outer orbit. Even for $e = 0.01$, $\mathcal{B} \sim 100$.}
    \label{fig: log10_BF}
\end{figure}

An inherent assumption of our analysis is that the correct model to be fit to the data is known. This will not be true in general. Moreover, use of the wrong model could lead to stealth biases. We show this in Figure \ref{fig: stealth_bias} of the Appendix~\ref{sec: supp_fig} where a $100-100\, {\rm M_{\odot}}$ binary at $1\, {\rm Gpc}$ in an eccentric outer orbit of semi-major axis $R = 5 \times 10^4 \, R_{\rm s}$ around a $10^6 \, {\rm M_{\odot}}$ SMBH, detected in DECIGO, is considered. For eccentricities (of the outer orbits) as low as $0.01$, fitting a circular orbit will result in a stealth bias of one of the outer orbit parameters. Nevertheless, it is possible to use Bayesian model selection first to determine which orbit is preferred: circular or eccentric. For this, we set $\cos \vartheta = -e$ in Eq. \ref{eq: ecc_orbit} to set $x = 1$ (i.e. $r = R$) so that we have only one extra parameter eccentricity $e$ in the true model. This allows us to use the Savage-Dickey density ratio to evaluate a Bayes Factor (see \cite{PhysRevD.99.124044} and the references therein). Following Eq. (5) of \cite{PhysRevD.99.124044}, we define the Bayes factor in favor of a non-zero eccentricity as: 
\begin{equation}
    \label{eq: Bayes_fact}
    \mathcal{B} = \sqrt{2 \pi} \sigma \exp{\frac{\mu_e^2}{2 \sigma^2 }} \frac{1}{\Delta \lambda}
\end{equation}
where $\mu_e$ is the median of the posterior of eccentricity, $\sigma$ is the standard deviation of the same, while $\Delta \lambda$ is the prior on the eccentricity which we take to be uniform between 0 and 1. As shown in Figure ~\ref{fig: log10_BF}, we find strong preferences for eccentric orbits, even for eccentricities as low as $e = 0.01$, with   Bayes factors increasing drastically for larger eccentricities.
    
To further test our inherent assumption of knowing the true model and the ability of our method to distinguish between different potentials, we consider the $30-30\, \rm M_{\odot}$ CBC in an eccentric orbit of eccentricity $e = 0.5$, semimajor axis $R = 2 \times 10^4\, R_{\rm s}$, and $x = 1$ around a 1$0^7 \rm M_{\odot}$ SMBH at $1\, \rm Gpc$ in DECIGO. We, as usual, compute the Fisher Matrix of the kinematic parameters: LOSA, LOSJ, LOSS, amd LOSC and then sample the parameters related to the outer orbit and potential. We first sample the parameters of the true model, i.e, eccentric orbit around an SMBH, and then use thermodynamic integration (see section 3 of \cite{10.1111/j.1467-9868.2007.00650.x} for the details of the method) to compute the evidence. We finally repeat the same process, assuming the CBC to be in a Bahcall-Wolf (BW) potential, the wrong model in this case. We find the log Bayes factor in favour of the eccentric orbit around an SMBH to be: $\ln \mathcal{B} = 195.2$. Clearly, the true model is decisively preferred over the wrong one. This suggests that we can distinguish between environment potentials using our method.

\section{Discussion}\label{sec:conclusion}
The astrophysical processes that govern compact binary coalescences (CBCs) are at best partially understood, if at all \cite{Spera:2022byb}. Several models of the environments in which CBCs form, evolve, and merge have been proposed \cite{Mapelli:2021taw}, although there is no conclusive evidence to suggest that one formation channel is preferred over others. Indeed, some works suggest that multiple formation channels have contributed to the production of the detected stellar-mass CBCs to date \cite{Zevin:2020gbd, Pierra:2024fbl}. The intrinsic parameters of a CBC, viz., its component masses, spins, and/or orbital eccentricity, can sometimes be suggestive of a subset of possible merger environments and formation channels \cite{Farr:2017uvj, Bavera:2020inc, Bogdanovic:2007hp, Vajpeyi:2021qsw, McKernan:2023xio, Adamcewicz:2023szp, Samsing:2020tda, Tagawa:2020jnc, Kozai:1962zz, Lidov:1962wjn, Naoz2016, Antonini:2017ash, Yu:2020iqj, Randall:2017jop, Gerosa:2021mno, Liu:2020gif}. They cannot, however, always conclusively rule out all other possible formation channels, especially if there is no EM counterpart. 

In this work, we show that rich and precise information of the environment with {\it generic} gravitational potentials can be extracted {\it on a single-event basis}. Additionally, we show that it is also possible to acquire information about the location of the CBC with respect to the center of the potential. This can be achieved provided the CBC's LOSA and some of its higher time derivatives can be measured. We consider the SMBH, NSC (BW), and GC (Plummer) potentials. We find that we can measure enclosed mass and potential profile exponents, and the eccentricity and the radius of the CBC's orbit in the potential, with relative errors of less than unity for a large portion of the parameter space we consider, even going down to sub-percent precision for certain portions.

An important assumption inherent in our analysis is that we know the true model (environment + outer-orbit) to fit to the data. This, of course, is not the case in general. Moreover, as we show in Figure~\ref{fig: stealth_bias} of the Appendix~\ref{sec: supp_fig}, fitting a wrong model to the data will result in potential stealth biases. Nevertheless, such biases can be avoided by selecting the correct model via Bayesian model selection. By evaluating the Savage-Dickey density ratio~\cite{PhysRevD.99.124044}, we demonstrate in Figure~\ref{fig: log10_BF} that it should be possible to differentiate the true orbit from other incorrect alternatives, given the same potential. We also perform a full Bayes factor analysis and show that two different potentials can also be confidently distinguished.

We expect that our work, demonstrating the possibility of constraining environment parameters, {\it on a single-event basis}, could have significant astrophysical implications, potentially answering the longstanding question of the provenance of CBCs for at least a fraction of them with measurable LOSV time derivatives (see \cite{Tiwari:2023cpa} for estimates of this fraction , focusing exclusively on LOSAs, for various GC configurations). Moreover, it could unearth important correlations between the mass of the CBC, the mass of the environment, the steepness of the potential profile, and the location of the CBC with respect to the center of the potential. Probing these would directly test astrophysical mechanisms, such as dynamical friction and mass segregation, among others, in dense stellar environments.

Strictly, our work probes the environment near the end of the CBC's lifetime. Given that the delay time between the CBC's formation and merger could span millions to billions of years (see, e.g., \cite{Fishbach:2021mhp}, and references therein), the location of the environment probed by the CBC could be quite different from the location of its formation. This is due to kicks induced by, e.g., asymmetric supernova mechanisms, or interactions with other bodies (the probability of which is relatively higher in dense star clusters~\cite{1987degc.book.....S, 2008gady.book.....B}). 

Thus, our work, complemented by observations of binary stars, could enable the construction of the complete story of their evolution, starting out as well-separated (bound) main-sequence stars, and ending with the coalescence of compact objects.

\begin{acknowledgments}
We thank Debatri Chattopadhyay for their careful reading and feedback. We thank K. G. Arun, Debarati Chatterjee and Zoheyr Doctor for useful discussions. SJK gratefully acknowledges support from SERB Grants SRG/2023/000419 and MTR/2023/000086. SC acknowledges support from the Department of Atomic Energy, Government of India, under project no. 12-R\&D-TFR-5.02-0200 and RTI 4002. AV acknowledges support to CITA by the Natural Sciences and Engineering Research Council of Canada (NSERC) (funding reference number 568580). All computations were performed on the Sarathi Computing Cluster at IUCAA.
\end{acknowledgments}

\textit{Software}: \texttt{NumPy} \citep{vanderWalt:2011bqk}, \texttt{SciPy} \citep{Virtanen:2019joe}, \texttt{astropy} \citep{2013A&A...558A..33A, 2018AJ....156..123A}, \texttt{Matplotlib} \citep{Hunter:2007}, \texttt{jupyter} \citep{jupyter}, \texttt{Numdifftools} \citep{numdifftools}.

\bibliography{apssamp}

\onecolumngrid
\appendix

\section{Gravitational-Wave Phase Correction due to time derivatives of the Line-of-Sight Velocity}
\label{sec: phase_cor}
We follow Ref.~\cite{Vijaykumar_2023} to derive the correction to the gravitational-wave (GW) phase incurred by finite time-derivatives of line-of-sight velocity (LOSV) of the center of mass (CoM) of a compact binary coalescence (CBC). We compute the leading Post Newtonian (PN) order phase correction due to the $n^{th}$ derivative of LOSV. We then acquire phase corrections due to specific kinematic parameters, viz. LOS acceleration (LOSA) $\Gamma_1$, jerk (LOSJ) $\Gamma_2$, snap (LOSS) $\Gamma_3$, crackle (LOSC) $\Gamma_4$, pop (LOSP) $\Gamma_5$, and lock (LOSL) $\Gamma_6$. 

\subsection{Phase Correction due to $n^{th}$ derivative of the LOSV}\label{subsec: phase_cor_n}
Let $z_{l,n} \equiv \Gamma_n (t_{\rm o} - t_{\rm c})^n$ be the (time-dependent) redshift due to the $n^{th}$ derivative of LOSV of the CoM of the binary having total mass $M$ and symmetric mass ratio $\eta$, evaluated at the coalescence time $t_{\rm c} $ in contrast to the time at which signal enters in the detector band in Ref.~\cite{Vijaykumar_2023} because it removes any possible coupling between $\Gamma_n$ and $t_{\rm c} $. Here $t_{\rm o} $ is the observation time, and $\Gamma_n \equiv \left. \frac{1}{n! c}\frac{d^n v_l}{dt_{\rm o} ^n} \right \vert_{t_{\rm o} = t_{\rm c}}$. Let $f_o$ and $f_u$ be the perturbed and unperturbed Fourier frequencies, respectively. Then under the assumption $| \Gamma_n (t_{\rm o} - t_{\rm c})^n | \ll 1$, equations (A1-A4) of \cite{Vijaykumar_2023} take the form:
\begin{align}
    \label{eq: f_uo_rel_n}
    f_u &= f_o (1 + z_{l,n}) \\
    \label{eq: dt_uo_rel_n}
    dt_u &= dt_{\rm o}  / (1 + z_{l,n}) \\
    \label{eq: v_uo_rel_n}
    v_u &= v_o (1 + z_{l,n})^{1/3} 
\end{align}
\begin{equation}
    \label{eq: dvdt_n}
    \frac{d v_o}{d t_{\rm o} } = - \frac{n}{3} \Gamma_n  v_o (t_{\rm o} - t_{\rm c})^{n-1} + \left[1 + \Gamma_n (t_{\rm o} - t_{\rm c})^n \right]^{-4/3} \frac{d v_u}{d t_u}
\end{equation}
where $v_u = (\pi M f_u)^{1/3}$ and $v_o = (\pi M f_o)^{1/3}$. We use geometrized units $G = c = 1$ unless otherwise mentioned.

Following the steps delineated in \cite{Vijaykumar_2023}, we get $\frac{dt_{\rm o} }{dv_o}$ and thereon get the (negative) time until coalescence $t(f) - t_{\rm c} $, orbital phase $\phi(f)$, and phase correction $\Delta \Psi_{-4n} (f)$ to the \texttt{TaylorF2} phase. The equations (A5-A7) of the same then take the form:
\begin{align}
    \label{eq: negcoal_n}
    t(f) - t_{\rm c}  &= -\frac{5 M }{256 \eta  v^8} \left[1-\frac{\Gamma_n  (8 n+5) v^{-8 n}}{3 (n+1)} \left(\frac{5}{256}\right)^n \left(-\frac{M}{\eta }\right)^n \right] \\
    \label{eq: phi_n}
    \phi(f) &= \frac{\phi_{\rm c} }{2} - \frac{1}{32 \eta  v^5} \left[ 1-\frac{1}{3} \Gamma_n  2^{-8 n} 5^{n+1} v^{-8 n} \left(-\frac{M}{\eta }\right)^n \right]
\end{align}
\begin{equation}
    \label{eq: phase_cor_n}
    \Delta \Psi_{-4n} (f) =  -\frac{ 2^{-8 n-7} 5^{n+1} v^{-8 n-5} \left(-\frac{M}{\eta }\right)^n}{\eta  (n+1)} \Gamma_n
\end{equation}
where $v \equiv (\pi M f)^{1/3}$ and $f$ is the observed frequency. The $n^{th}$ derivative of the LOSV introduces a $-4n$ PN correction to the GW phase. To obtain this leading order phase correction, i.e., equation~\ref{eq: phase_cor_n}, only quadrupole order terms in the phase and time to coalescence were required as opposed to 3.5PN NLO (next to leading order) terms in \cite{Vijaykumar_2023}. 

\subsection{Phase Corrections up to $-24 \mathrm{PN}$ order}\label{subsec: phase_cor_5}
The leading order phase correction due to a particular kinematic parameter will have contributions from higher powers of kinematic parameters pertaining to lower time derivatives of LOSV. For example, the $-8$PN phase correction $\Delta \Psi_{-8} (f)$ will have contributions from a constant line of sight jerk (LOSJ) $\Gamma_2$ as well as the LOSA term proportional to $\Gamma_1^2$. We, therefore, calculate the phase corrections up to $-24$PN order considering contributions from derivatives up to LOSL.

Let us define $z_{l6} = \sum_{n = 1}^6 z_{l,n}$. The Eq.~\eqref{eq: f_uo_rel_n}--Eq.~\eqref{eq: v_uo_rel_n} then retain the same form with $z_{l,n}$ replaced by $z_{l6}$ while equations Eq.~\eqref{eq: dvdt_n}--Eq.\eqref{eq: phi_n} become:
\begin{equation}
    \label{eq: dvdt_5}
    \frac{dv_o}{dt_{\rm o} } = - \frac{v_o}{3} \frac{\sum_{n = 1}^6 n \Gamma_n (t_{\rm o} - t_{\rm c})^{n-1}}{1 + \sum_{n = 1}^6 \Gamma_n (t_{\rm o} - t_{\rm c})^n} + \left[ 1 + \sum_{n = 1}^6 \Gamma_n (t_{\rm o} - t_{\rm c})^n \right]^{-4/3} \frac{dv_u}{dt_u}
\end{equation}
\begin{multline}
    \label{eq: negcoal_5}
    t(f) - t_{\rm c}  = -\frac{5 M}{256 \eta  v^8} \Biggl[ 1 + \frac{65 \Gamma_1 M}{1536 \eta  v^8} + \frac{M^2}{\eta ^2 v^{16}} \Biggl( \frac{1175 \Gamma_1^2}{442368}-\frac{175 \Gamma_2}{196608} \Biggr) + \frac{M^3}{\eta ^3 v^{24}} \Biggl(\frac{258875 \Gamma_1^3}{1358954496} -\frac{2375 \Gamma_1 \Gamma_2}{18874368} +\frac{3625 \Gamma_3}{201326592} \Biggr) \\ + \frac{M^4}{\eta ^4 v^{32}} \Biggl( \frac{2537375 \Gamma_1^4}{173946175488}-\frac{416875 \Gamma_1^2 \Gamma_2}{28991029248} +\frac{4375 \Gamma_1 \Gamma_3}{1610612736} +\frac{15125 \Gamma_2^2}{9663676416}-\frac{4625 \Gamma_4}{12884901888} \Biggr) \\ +  \frac{M^5}{\eta ^5 v^{40}} \Biggl( \frac{932646875 \Gamma_1^5}{801543976648704} -\frac{305678125 \Gamma_1^3 \Gamma_2}{200385994162176}+\frac{1596875 \Gamma_1^2 \Gamma_3}{4947802324992}+\frac{615625 \Gamma_1 \Gamma_2^2}{1649267441664}-\frac{209375 \Gamma_1 \Gamma_4}{3710851743744} \\ -\frac{259375 \Gamma_2 \Gamma_3}{3710851743744}+\frac{15625 \Gamma_5}{2199023255552} \Biggr) + \frac{M^6}{\eta ^6 v^{48}} \Biggl( \frac{308525140625 \Gamma_1^6}{3231825313847574528} - \frac{112183765625 \Gamma_1^4 \Gamma_2}{718183403077238784} \\ + \frac{2438265625 \Gamma_1^2 \Gamma_2^2}{39899077948735488} - \frac{18765625 \Gamma_2^3}{5699868278390784} + \frac{2105265625 \Gamma_1^3 \Gamma_3}{59848616923103232} - \frac{114109375 \Gamma_1 \Gamma_2 \Gamma_3}{6649846324789248} \\ + \frac{3546875 \Gamma_3^2}{4433230883192832} - \frac{91609375 \Gamma_1^2 \Gamma_4}{13299692649578496} + \frac{3296875 \Gamma_2 \Gamma_4}{2216615441596416} + \frac{2546875 \Gamma_1 \Gamma_5}{2216615441596416} - \frac{828125 \Gamma_6}{5910974510923776} \Biggr) \Biggr]
\end{multline}
\begin{multline}
    \label{eq: phi_5}
    \phi (f) = \frac{\phi_{\rm c} }{2} - \frac{1}{32 v^5 \eta} \Biggl[1 +  \frac{25 \Gamma_1 M}{768 \eta  v^8} + \frac{M^2}{\eta ^2 v^{16}} \left(\frac{5875 \Gamma_1^2}{3096576}-\frac{125 \Gamma_2}{196608}\right) + \frac{M^3}{\eta ^3 v^{24}} \Biggl(\frac{1294375 \Gamma_1^3}{9852420096} -\frac{11875 \Gamma_1 \Gamma_2}{136839168}+\frac{625 \Gamma_3}{50331648}\Biggr) \\ + \frac{M^4}{\eta ^4 v^{32}} \Biggl( \frac{63434375 \Gamma_1^4}{6436008493056}-\frac{10421875 \Gamma_1^2 \Gamma_2}{1072668082176} +\frac{109375 \Gamma_1 \Gamma_3}{59592671232}+\frac{378125 \Gamma_2^2}{357556027392}-\frac{3125 \Gamma_4}{12884901888} \Biggr) \\ + \frac{M^5}{\eta ^5 v^{40}} \Biggl( \frac{932646875\Gamma_1^5}{1202315964973056} -\frac{305678125 \Gamma_1^3 \Gamma_2}{300578991243264}+\frac{1596875 \Gamma_1^2 \Gamma_3}{7421703487488}  +\frac{615625 \Gamma_1 \Gamma_2^2}{2473901162496} -\frac{209375 \Gamma_1 \Gamma_4}{5566277615616} \\ -\frac{259375 \Gamma_2 \Gamma_3}{5566277615616}+\frac{15625 \Gamma_5}{3298534883328} \Biggr) + \frac{M^6}{\eta ^6 v^{48}} \Biggl( \frac{1542625703125 \Gamma_1^6}{24469534519131635712} -\frac{560918828125 \Gamma_1^4 \Gamma_2}{5437674337584807936} +\frac{12191328125 \Gamma_1^2 \Gamma_2^2}{302093018754711552} \\ -\frac{656796875 \Gamma_2^3}{302093018754711552} +\frac{10526328125 \Gamma_1^3 \Gamma_3}{453139528132067328} -\frac{570546875 \Gamma_1 \Gamma_2 \Gamma_3}{50348836459118592} +\frac{17734375 \Gamma_3^2}{33565890972745728} \\ -\frac{458046875 \Gamma_1^2 \Gamma_4}{100697672918237184} +\frac{16484375 \Gamma_2 \Gamma_4}{16782945486372864} +\frac{12734375 \Gamma_1 \Gamma_5}{16782945486372864} -\frac{78125 \Gamma_6}{844424930131968} \Biggr) \Biggr]
\end{multline}

The different order phase corrections are then given by:
\begin{equation}
    \label{eq: app_pcj}
    \Delta \Psi_{-8} (f) = \frac{M^2}{\eta^3 v^{21}} \Biggl[ - \frac{125 \Gamma_2}{25165824} + \frac{5875 \Gamma_1^2}{396361728} \biggr]
\end{equation}
\begin{equation}
    \label{eq: app_pcs}
    \Delta \Psi_{-12} (f) = \frac{M^3}{\eta^4 v^{29}} \Biggl[ \frac{625  \Gamma_3}{8589934592} + \frac{1294375 \Gamma_1^3}{1681479696384}-\frac{11875 \Gamma_1 \Gamma_2}{23353884672} \Biggr]
\end{equation}
\begin{multline}
    \label{eq: app_pcc}
    \Delta \Psi_{-16} (f) =  \frac{M^4}{\eta^5 v^{37}} \Biggl[-\frac{625 \Gamma_4}{549755813888} + \frac{12686875 \Gamma_1^4}{274603029037056} -\frac{2084375 \Gamma_1^2 \Gamma_2}{45767171506176}  +\frac{21875 \Gamma_1 \Gamma_3}{2542620639232}+\frac{75625 \Gamma_2^2}{15255723835392} \Biggr]
\end{multline}
\begin{multline}
    \label{eq: app_pcp}
    \Delta \Psi_{-20} (f) = \frac{M^5}{\eta^6 v^{45}} \Biggl[\frac{15625 \Gamma_5}{844424930131968} + \frac{932646875 \Gamma_1^5}{307792887033102336} -\frac{305678125 \Gamma_1^3 \Gamma_2}{76948221758275584} \\ +\frac{1596875 \Gamma_1^2 \Gamma_3}{1899956092796928} +\frac{615625 \Gamma_1 \Gamma_2^2}{633318697598976}  -\frac{209375 \Gamma_1 \Gamma_4}{1424967069597696} -\frac{259375 \Gamma_2 \Gamma_3}{1424967069597696} \Biggr]
\end{multline}
\begin{multline}
    \label{eq: app_pcl}
    \Delta \Psi_{-24} (f) = \frac{M^6}{\eta^7 v^{53}}\Biggl[- \frac{78125 \Gamma_6}{252201579132747776} + \frac{1542625703125 \Gamma_1^6}{7308234309713981865984}-\frac{560918828125 \Gamma_1^4 \Gamma_2}{1624052068825329303552} \\ + \frac{10526328125 \Gamma_1^3 \Gamma_3}{135337672402110775296}+\frac{12191328125 \Gamma_1^2 \Gamma_2^2}{90225114934740516864}-\frac{458046875 \Gamma_1^2 \Gamma_4}{30075038311580172288} - \frac{570546875 \Gamma_1 \Gamma_2 \Gamma_3}{15037519155790086144} \\ + \frac{12734375 \Gamma_1 \Gamma_5}{5012506385263362048}  - \frac{93828125 \Gamma_2^3}{12889302133534359552} + \frac{16484375 \Gamma_2 \Gamma_4}{5012506385263362048} + \frac{17734375 \Gamma_3^2}{10025012770526724096} \Biggr]
\end{multline}
We have not written $\Delta \Psi_{-4}$ here as this is given by Eq. (A7) of \cite{Vijaykumar_2023} with $\Gamma$ replaced by $\Gamma_1$. 
Moreover, we calculate the next-to-leading order corrections due to LOSJ following again the steps of \cite{Vijaykumar_2023}. The final expression for $\Delta \Psi_{-8} (f)$ then takes the form:
\begin{multline}
    \label{eq: pcj_nlo}
    \Delta \Psi_{-8} (f) = - \frac{125 \Gamma_2  M^2}{25165824 \eta ^3 v^{21}} \Biggl[ 1 + \left(11 \eta + \frac{743}{84} \right) v^2 - \frac{96 \pi}{5}v^3 + \Biggl( \frac{1585 \eta ^2}{24}+\frac{6997 \eta }{72} +\frac{7475065}{169344} \Biggr) v^4  \\ - \left( \frac{639\eta}{5} + \frac{86197 }{420} \right) \pi v^5  + \Biggl\{ \frac{6848}{35}(\gamma + \log(4v)) + \frac{128305 \eta ^3}{432} + \frac{1939751 \eta ^2}{4032} + \left(\frac{3572915219}{1016064} - \frac{451 \pi ^2}{4} \right) \eta \\ + \frac{6272 \pi ^2}{25}  - \frac{81167909790719}{70413235200}  \Biggr\} v^6 - \left( \frac{235831 \eta^2}{630} + \frac{241819 \eta }{140} + \frac{92073173}{70560} \right) \pi v^7  \Biggr] + \frac{5875 \Gamma_1^2 M^2}{396361728 \eta^3 v^{21}}
\end{multline}

This is the expression of $\Delta \Psi_{-8} (f)$ that we use in the Fisher matrix calculations. Corrections to the phase due to higher time derivatives of LOSV are evaluated only at leading PN order.

\section{Amplitude correction}\label{sec: amp_cor}
Even though we do not use the amplitude corrections in the Fisher Matrix analysis, we mention the results for completeness. Similar to the previous section, following the steps of \cite{Vijaykumar_2023}, we get the leading order relative amplitude correction due to $n^{th}$ derivative of the LOS velocity as:
\begin{equation}
    \label{eq: ac_n}
    \frac{\Delta \mathcal{A}_{-4n}}{\mathcal{A}} = -  \frac{5^n \Gamma_n}{3 (n+1) v^{8 n}} \left(-\frac{M}{\eta }\right)^n \left[ n^2 \cdot 4^{1-4 n}  +  9n\cdot 2^{-8 n-1}  +  5 \cdot 4^{-4 n-1} \right]
\end{equation}
Note, however, that these too will have contributions from higher powers of lower-order derivatives of the LOS velocity. Hence using the definition of $z_{l6}$ and following the steps of \cite{Vijaykumar_2023}, we get the different order relative amplitude corrections as:
\begin{equation}
    \label{eq: amp_cor_8}
    \frac{\Delta \mathcal{A}_{-8}}{\mathcal{A}} = \frac{M^2}{\eta^2 v^{16}} \Biggl[\frac{511175 \Gamma_1^2}{226492416}-\frac{875\Gamma_2}{786432} \Biggr]
\end{equation}
\begin{equation}
    \label{ep: amp_cor_12}
    \frac{\Delta \mathcal{A}_{-12}}{\mathcal{A}} = \frac{M^3}{\eta^3 v^{24}} \Biggl[ \frac{250429375 \Gamma_1^3}{1391569403904}-\frac{734125 \Gamma_1 \Gamma_2}{4831838208}+\frac{25375 \Gamma_3}{805306368} \Biggr]
\end{equation}
\begin{multline}
    \label{ep: amp_cor_16}
    \frac{\Delta \mathcal{A}_{-16}}{\mathcal{A}} = \frac{M^4}{\eta^4 v^{32}} \Biggl[ \frac{1536363228875 \Gamma_1^4}{102597629011034112}-\frac{3155433125\Gamma_1^2 \Gamma_2}{178120883699712}+\frac{7016875\Gamma_1 \Gamma_3}{1649267441664} +\frac{3039125\Gamma_2^2}{1236950581248}-\frac{13875 \Gamma_4}{17179869184} \Biggr]
\end{multline}
\begin{multline}
    \label{ep: amp_cor_20}
    \frac{\Delta \mathcal{A}_{-20}}{\mathcal{A}} = \frac{M^5}{\eta^5 v^{40}} \Biggl[ \frac{807473283848125  \Gamma_1^5}{630359832643793584128}-\frac{6372397146875  \Gamma_1^3  \Gamma_2}{3283124128353091584}  +\frac{9947365625 \Gamma_1^2  \Gamma_3}{20266198323167232} \\+\frac{4324094375 \Gamma_1  \Gamma_2^2}{7599824371187712}-\frac{103208125  \Gamma_1  \Gamma_4}{949978046398464}  -\frac{256721875  \Gamma_2  \Gamma_3}{1899956092796928}+\frac{171875 \Gamma_5}{8796093022208} \Biggr]
\end{multline}
\begin{multline}
    \label{ep: amp_cor_24}
    \frac{\Delta \mathcal{A}_{-24}}{\mathcal{A}} = \frac{M^6}{\eta^6 v^{48}} \Biggl[\frac{6056019825399034375 \Gamma_1^6}{54221031364688548932354048} - \frac{116358888203309375 \Gamma_1^4 \Gamma_2 }{564802410048839051378688} + \frac{139779250765625 \Gamma_1^3 \Gamma_3}{2614825972448328941568} \\ + \frac{182461202903125 \Gamma_1^2 \Gamma_2^2}{1961119479336246706176}-\frac{1018926090625 \Gamma_1^2 \Gamma_4}{81713311639010279424} - \frac{2549078359375 \Gamma_1 \Gamma_2 \Gamma_3}{81713311639010279424} + \frac{2979640625 \Gamma_1 \Gamma_5}{1134907106097364992} \\ - \frac{17477759375 \Gamma_2^3}{2918332558536081408} + \frac{969240625 \Gamma_2 \Gamma_4}{283726776524341248} +\frac{16709171875 \Gamma_3^2}{9079256848778919936}-\frac{10765625 \Gamma_6}{23643898043695104} \Biggr]
\end{multline}

As can be seen from the {\it right panel} of Figure \ref{fig: pc_ac_BW}, the relative amplitude corrections due to all derivatives of LOSV are $\ll 1$ with the largest contribution coming from the LOSA. We have checked that including the amplitude corrections due to LOSA in the Fisher Matrix does not affect the results. This is why we do not include the amplitude corrections in the Fisher Matrix.

\section{Kinematic parameters for various orbits in different gravitational potentials}\label{sec: kin_param_derivation}

We derive the relationship between kinematic parameters and environment parameters for a set of orbits, including Circular orbit, Elliptical orbit, and Effective Keplerian orbit, in different gravitational potentials. These potentials are: Keplerian (valid in the vicinity of a supermassive black hole (SMBH)), Bahcall-Wolf (valid for nuclear star clusters (NSCs) \cite{Hoang_2018}), and Plummer (valid for globular clusters (GCs) \cite{Plummer1911}). 

\subsection{Circular Orbit}\label{subsec: app_CO}
Consider a binary of total mass $M$ orbiting in a circular orbit of radius $R$. Let $M_{\mathrm{enc}}$ be the enclosed mass within this radius. Then the acceleration of the CoM of the binary in the reference frame mentioned above is given by:
\begin{equation}
    \label{eq: acc_CO}
    \boldsymbol{a} = - \frac{G M_{\mathrm{enc}}}{R^2} \boldsymbol{e}_r = - \omega^2 R \boldsymbol{e}_r
\end{equation}
where $\omega \equiv \Dot{\vartheta} = \sqrt{\frac{G M_{\mathrm{enc}}}{R^3}}$ is the angular frequency of the CoM's outer orbit. Since for circular orbits $\Dot{r} = 0$, using this together with $\omega$ and replacing $\boldsymbol{\mathcal{V}}$ in Eq. \eqref{eq: vectder} by $\boldsymbol{a}$ gives us jerk ($\boldsymbol{j}$). Similarly, replacing the same with jerk gives us snap ($\boldsymbol{s}$):
\begin{align}
    \label{eq: jerk_CO}
    \boldsymbol{j} &= - \omega^3 R \boldsymbol{e}_{\vartheta} \\
    \label{eq: snap_CO}
    \boldsymbol{s} &=  \omega^4 R \boldsymbol{e}_r
\end{align}
In particular, a $n^{th}$ order derivative of the velocity will vary as $\sim \omega^{n+1} R$. Projecting these quantities along our LOS $\boldsymbol{n}$ and using equations \eqref{eq: unitvect}, we get:
\begin{align}
    \label{eq: los_acc_CO}
    a &= - \omega^2 R \sin \vartheta \sin \iota_{\mathrm{out}} \\
    \label{eq: los_jerk_CO}
    j &= - \omega^3 R \cos \vartheta \sin \iota_{\mathrm{out}} \\
    \label{eq: los_snap_CO}
    s &= \omega^4 R \sin \vartheta \sin \iota_{\mathrm{out}}
\end{align}
These equations indicate that we would not be able to measure $\sin \iota_{\mathrm{out}}$ because it will always be degenerate with $R$ and $M_{\mathrm{enc}}$. After some manipulation, we get the inverse relations as follows:
\begin{equation}
    \label{eq: inv_rel_CO}
    \begin{aligned}
        \cos \vartheta &= \pm \left(1 - \frac{a \cdot s}{j^2} \right)^{-1/2} \\ 
        \omega &= \sqrt{- \frac{s}{a}} \\
        R \sin \iota_{\mathrm{out}} &= \pm \frac{a^2}{s} \left(\frac{j^2}{a \cdot s} - 1 \right)^{1/2} \\
        M_{\mathrm{enc}} \sin^3 \iota_{\mathrm{out}} &= \mp \frac{1}{G}\frac{a^5}{s^2} \left(\frac{j^2}{a \cdot s} - 1 \right)^{3/2}
    \end{aligned}
\end{equation}
It should be noted that even though we have derived the inverse relations, we use them only to confirm the results from the Markov Chain Monte Carlo(MCMC) sampling of the GW likelihood (see the Sec.~\ref{subsec: fma}) expressed as a function of $M_{\mathrm{SMBH}}$, $R$, and $\cos \vartheta$, using equations \eqref{eq: los_acc_CO} - \eqref{eq: los_snap_CO}.

Due to a constant radius, the binary in circular orbit does not trace the variation in gravitational potential and hence density profile. Therefore, all that we get is enclosed mass, and we can not distinguish between different sources of gravitational potential and hence, without loss of generality, we consider a circular orbit around an SMBH in this work. The analysis would be identical for any spherically symmetric potential, with $M_{\mathrm{SMBH}}$ replaced by $M_{\mathrm{enc}}$.

\subsection{Eccentric orbit around an SMBH}\label{subsec: app_EO}
Consider a binary orbiting an SMBH of mass $M_{\mathrm{SMBH}}$ in an elliptical orbit of eccentricity $e$ and semi-major axis $R$. In the frame mentioned above with $\boldsymbol{e}_x$ pointing towards periapsis, we can write the Eq. of the orbit as:
\begin{equation}
    \label{eq: ecc_orbit}
    r = \frac{R (1 - e^2)}{1 + e \cos \vartheta}
\end{equation}
A rearrangement of the terms in the above Eq. gives us:
\begin{align}
    \label{eq: cos_EO}
    \cos \vartheta &= \frac{x (1 - e^2) - 1}{e} \\
    \label{eq: sin_EO}
    \sin \vartheta &= \frac{\sqrt{1 - e^2}}{e}\sqrt{2x - (1 - e^2)x^2 - 1}
\end{align}
where $x \equiv \frac{R}{r}$ with $r$ as the distance of the merger from the SMBH. Using the fact that the total energy $E_{\mathrm{out}}$ and angular momentum $L_{\mathrm{out}}$, per unit mass, are given by $- \frac{G M_{\mathrm{SMBH}}}{2 R}$ and $\sqrt{G M_{\mathrm{SMBH}} R (1 - e^2)}$, respectively, we can write $\Dot{r}$ and $\Dot{\vartheta}$ as: 
\begin{align}
    \label{eq: rdot_EO}
    \Dot{r} &= R \omega  \sqrt{e^2 x^2-\left(1-x\right)^2} \\
    \label{eq: thdot_EO}
    \Dot{\vartheta} &= \sqrt{1-e^2} \omega x^2
\end{align}
where $\omega = \sqrt{\frac{G M_{\mathrm{SMBH}}}{R^3}}$ is the orbital angular frequency averaged over one orbital period.

The acceleration\footnote{Since the acceleration in a gravitational field is always towards the centre, from hereon we will refer to the radial component of acceleration as the acceleration for simplicity.}, in this case, is given by:
\begin{equation}
    \label{eq: acc_EO}
    a_r = - \frac{G M_{\mathrm{SMBH}}}{r^2} = - \omega^2 R x^2
\end{equation}
Similar to the circular orbit case, replacing $\boldsymbol{\mathcal{V}}$ in Eq. \eqref{eq: vectder} by $\boldsymbol{a}$ and utilising equations \eqref{eq: rdot_EO} and \eqref{eq: thdot_EO} gives us $\boldsymbol{j}$ while replacing the same with $\boldsymbol{j}$ gives us $\boldsymbol{s}$ and so on. The radial and angular components of $\boldsymbol{j, s, cr}$ are then given by:
\begin{itemize}
    \item Jerk ($\boldsymbol{j}$):
    \begin{equation}
        \begin{aligned}
            \label{eq: jerk_EO}
            j_r &= 2 \omega^3 R x^3 \sqrt{e^2 x^2-(1-x)^2} \\
            j_{\vartheta} &= -\omega^3 R x^4 \sqrt{1-e^2}
        \end{aligned}
    \end{equation}
    \item Snap ($\boldsymbol{s}$):
    \begin{equation}
        \begin{aligned}
            \label{eq: snap_EO}
            s_r &= \omega^4 R x^4 \left[9 \left(1-e^2\right) x^2-14 x+6\right] \\
            s_{\vartheta} &= 6 \omega^4 R x^5 \sqrt{1-e^2} \sqrt{2 x -1 - \left(1-e^2\right) x^2}
        \end{aligned}
    \end{equation}
    \item Crackle ($\boldsymbol{cr}$):
    \begin{equation}
        \begin{aligned}
            \label{eq: crac_EO}
            cr_r &= -2 \omega^5  R x^5 \left[30 \left(1-e^2\right) x^2-35 x+12\right] \sqrt{2 x - 1 - \left(1-e^2\right) x^2} \\
            cr_{\vartheta} &= -\omega ^5 R x^6 \left[-45 \left(1-e^2\right) x^2+80 x-36\right] \sqrt{1-e^2}
        \end{aligned}
    \end{equation}
\end{itemize}
Projecting these quantities along our LOS $\boldsymbol{n}$ and using equations \eqref{eq: cos_EO} and \eqref{eq: sin_EO} gives us LOSA, LOSJ, LOSS, and LOSC:
\begin{align}
    \label{eq: los_acc_EO}
    a &= - x^2 \omega^2 R \sqrt{2 x- (1 - e^2) x^2 - 1} \frac{\sqrt{1-e^2}}{e} \sin \iota_{\mathrm{out}} \\
    \label{eq: los_jerk_EO}
    j &= x^3 \omega^3 R [5 x - 3 (1 - e^2) x^2 - 2] \frac{ \sqrt{1-e^2} }{e} \sin \iota_{\mathrm{out}} \\
    \label{eq: los_snap_EO}
    s &= - x^4 \omega ^4 R \sqrt{2 x- (1 - e^2) x^2 - 1} [20 x - 15 (1 - e^2) x^2 - 6] \frac{ \sqrt{1-e^2} }{e} \sin \iota_{\mathrm{out}} \\
    \label{eq: los_crac_EO}
    cr &= x^5 \omega^5 R  [105 (1 - e^2)^2 x^4 - 315 (1 - e^2) x^3 + 20 (17 - 6 e^2) x^2 - 154 x + 24] \frac{\sqrt{1-e^2}}{e} \sin \iota_{\mathrm{out}}
\end{align}
respectively. Here, we have fixed the projection of $\boldsymbol{n}$ in the outer orbital plane to be along the Y-axis. However, in elliptical orbits, this is not strictly general, as we do not have this freedom once we fix the X-axis to be pointing along the periapsis. In such a scenario, $\boldsymbol{n}$ should be taken as $(\cos \vartheta_{n} \boldsymbol{e}_x + \sin \vartheta_{n} \boldsymbol{e}_y) \sin \iota_{\mathrm{out}}$ where $\vartheta_n$ is the angle between X-axis and the projection of $\boldsymbol{n}$ onto the (outer) orbital plane. In other words, in this work, we set $\vartheta_{n} = \pi/2$ for simplicity, although it should be possible (but more involved) to extend the analysis to a more generally oriented ellipse.

The inverse relations are given by:
\begin{equation}
    \label{eq: inv_rel_EO}
    \begin{aligned}
        R \sin \iota_{\mathrm{out}} &= - \frac{a}{x^2 \omega^2}\frac{e}{\sqrt{1-e^2} \sqrt{2 x- (1 - e^2) x^2 - 1}} \\
        \omega &= - \frac{j}{a x} \frac{\sqrt{2 x- (1 - e^2) x^2 - 1}}{5 x - 3 (1 - e^2) x^2 - 2} \\
        M_{\mathrm{SMBH}} \sin^3 \iota_{\mathrm{out}} &= \frac{R^3 \omega ^2}{G} \\
        e &= \sqrt{1 + \frac{10 x (5 - 3 z_1) + 3 (4 z_1 - 7) - \sqrt{20 x^2 (5 - 3 z_1) - 36 x (5 - 3 z_1) + 48 z_1 + 81}}{6 x^2 (3 z_1 - 5)}}
    \end{aligned}
\end{equation}
The value of $x$ is given by one of the solutions of the following Eq.:
\begin{equation}
    \label{eq: xsol_EO}
    c_0 + c_1 x+ c_2 x^2 + c_3 x^3 = 0
\end{equation}
where:
\begin{equation}
    \begin{aligned}
        c_0 &= -12 [196 z_1^2 + 12 z_1 (4 z_2^2 - 39 z_2 + 15) + 405 (z_2-1)] \\
        c_1 &= 12 [756 z_1^2 + 5 z_1 (36 z_2^2 - 357 z_2 + 133) + 1575 (z_2 - 1) ] \\
        c_2 &= - 60 [192 z_1^2 + 15 z_1 (3 z_2^2 - 30 z_2 + 11) + 400 (z_2 - 1)] \\
        c_3 &= 25 [ 192 z_1^2 + 15 z_1 (3 z_2^2 - 30 z_2 + 11) + 400 (z_2 - 1) ]
    \end{aligned}
\end{equation}
and:
\begin{equation*}
    \begin{aligned}
        z_1 &= \frac{a \cdot s}{j^2} \\
        z_2 &= \frac{cr \cdot a}{s \cdot j}
    \end{aligned}
\end{equation*}
As before, we use these relations only to confirm the results from MCMC sampling of the GW likelihood (see the Sec.~\ref{subsec: fma}) expressed as a function of $M_{\mathrm{SMBH}}$, $R$, $x$, and $e$ using equations \eqref{eq: los_acc_EO} - \eqref{eq: los_crac_EO}.

\subsection{Orbits in Bahcall-Wolf potential}\label{subsec: app_BW}

For binary mergers in NSCs having a massive black hole (MBH) at the center, we use the Bahcall-Wolf (BW) profile defined in Eq. (4) of \cite{Hoang_2018}. For convenience, we rewrite the mass density of BW as:
\begin{equation}
    \label{eq: mass_den_BW}
    \rho (r) = \frac{3-\alpha}{4\pi}k M_{\mathrm{MBH}}^{\frac{\alpha - 1}{2}} r^{-\alpha}
\end{equation}
where $M_{\mathrm{MBH}}$ is the mass of the MBH at the center of the cluster, $\alpha$ is the steepness parameter, and $k = 2 \left( \frac{G \sqrt{M_0}}{\sigma_0^2} \right)^{\alpha-3}$ with $M_0 = 3 \times 10^8 {\rm M}_{\odot}$ and $\sigma_0 = 200 \mathrm{km}/\mathrm{s}$ being constants. 

The gravitational potential, in this case, will have a collective contribution from the MBH and the above profile. Hence, the enclosed mass and gravitational potential are given by:
\begin{align}
    \label{eq: menc_BW}
    M_{\mathrm{enc}} &= M_{\mathrm{MBH}} + k M_{\mathrm{MBH}}^{\frac{\alpha - 1}{2}} r^{3-\alpha} \\
    \label{eq: pot_BW}
    \Phi (r) &= \frac{G k M_{\mathrm{MBH}}^{\frac{\alpha - 1}{2}} r^{2-\alpha }}{2-\alpha } - \frac{G M_{\mathrm{MBH}}}{r},\qquad \alpha \neq 2
\end{align}

Observe that the above potential is a combination of a power law and Kepler potential (inverse-square force law), which means the orbits for any random value of $\alpha$ will not necessarily be closed (unless the orbit is circular or $\alpha = 3$) which means defining quantities like eccentricity is not trivial in this case. Therefore, we stick with the radial velocity $v_r \equiv \Dot{r}$ and angular speed $\Omega \equiv \Dot{\vartheta}$ for calculating the derivatives of LOS velocity. Additionally, though for simplicity we do not consider the $\alpha = 2$ case here, the potential will be a combination of a logarithmic term $\propto \ln r$ and the Kepler potential, and the same exercise can be repeated to get the acceleration and its higher derivatives.

Let $E_{\mathrm{out}}$ and $L_{\mathrm{out}}$ be the specific total energy and angular momentum of the binary CoM's outer orbit. Then we can write $\Omega$ and $v_r$ as:
\begin{align}
    \label{eq: om_BW}
    \Omega &= \frac{L_{\mathrm{out}}}{r^2} \\
    \label{eq: vr_BW}
    v_r &= \sqrt{2 (E_{\mathrm{out}} - \Phi (r) )-r^2 \Omega ^2}
\end{align}
respectively. Differentiating the above equations with respect to $r$ then gives us:
\begin{align}
    \label{eq: dvbydr_BW}
    v'_r &= \frac{a_r + r \Omega ^2}{v_r} \\
    \label{eq: dombydr_BW}
    \Omega' &= -\frac{2 \Omega }{r}
\end{align}
where $a_r = - \Phi'(r)$ is the acceleration which takes the form:
\begin{equation}
    \label{eq: acc_BW}
    a_r = -G \left[\frac{M_{\mathrm{MBH}}}{r^2} + k M_{\mathrm{MBH}}^{\frac{\alpha - 1}{2}} r^{1-\alpha } \right]
\end{equation}

Just like in the previous two subsections, we follow the recursive process of calculating higher derivatives by utilizing equations \eqref{eq: dvbydr_BW} - \eqref{eq: acc_BW} to get $\boldsymbol{j, s, cr, p, l}$. While calculating the derivatives higher than the jerk, the higher derivatives of $v_r$ and $\Omega$ with respect to $r$ will also appear but those can again be re-written in terms of $r$, $v_r$, $\Omega$, and $a_r$ just like $v'_r$ and $\Omega'$ (see Eq. \eqref{eq: dvbydr_BW} and \eqref{eq: dombydr_BW}). The radial and angular components of these quantities are then given by\footnote{Here and in the next subsection, we write the radial and angular parts separately rather than the projected equations along our LOS to avoid the already cumbersome equations from being more cumbersome. It is always possible to project these along our LOS once we have both components.}:
\begin{itemize}
    \item Jerk ($\boldsymbol{j}$):
    \begin{equation}
        \label{eq: jerk_BW}
        \begin{aligned}
            j_r &= G v_r \left[\frac{2 M_{\mathrm{MBH}}}{r^3}-(1-\alpha ) k M_{\mathrm{MBH}}^{\frac{\alpha -1}{2}} r^{-\alpha }\right] \\
            j_{\vartheta} &= - G \Omega  \left[\frac{M_{\mathrm{MBH}}}{r^2} + k M_{\mathrm{MBH}}^{\frac{\alpha -1}{2}} r^{1-\alpha } \right]
        \end{aligned}
    \end{equation}
    \item Snap ($\boldsymbol{s}$):
    \begin{equation}
        \label{eq: snap_BW}
        \begin{aligned}
            s_r &= - G r^{-2 \alpha -5} \Biggl[G \left\{(\alpha -1) k^2 r^6 M_{\mathrm{MBH}}^{\alpha - 1}+(\alpha +1) k M_{\mathrm{MBH}}^{\frac{\alpha + 1}{2}} r^{\alpha +3}+2 M_{\mathrm{MBH}}^2 r^{2 \alpha }\right\} \\ & \quad +\alpha  k M_{\mathrm{MBH}}^{\frac{\alpha - 1}{2}} r^{\alpha +4} \left\{(\alpha -1) v_r^2-r^2 \Omega^2\right\} - 3 M_{\mathrm{MBH}} r^{2 \alpha +1} \left(r^2 \Omega ^2-2 v_r^2\right)\Biggr] \\
            s_{\vartheta} &= 2 G v_r \Omega  \left[\frac{3 M_{\mathrm{MBH}}}{r^3} + \alpha  k M_{\mathrm{MBH}}^{\frac{\alpha -1}{2}} r^{-\alpha }\right]
        \end{aligned}
    \end{equation}
    \item Crackle ($\boldsymbol{cr}$):
    \begin{equation}
        \label{eq: crackle_BW}
        \begin{aligned}
            cr_r &= G v_r r^{-2 (\alpha +3)} \Biggl[G \biggl\{\left(4 \alpha ^2-5 \alpha +1\right) k^2 r^6 M_{\mathrm{MBH}}^{\alpha - 1}+\left(3 \alpha ^2+\alpha +14\right) k M_{\mathrm{MBH}}^{\frac{\alpha +1}{2}} r^{\alpha +3}  +22 M_{\mathrm{MBH}}^2 r^{2 \alpha }\biggr\} \\ & \quad +\alpha  (\alpha +1) k M_{\mathrm{MBH}}^{\frac{\alpha - 1}{2}} r^{\alpha +4} \left\{(\alpha -1) v_r^2-3 r^2 \Omega ^2\right\} +12 M_{\mathrm{MBH}} r^{2 \alpha +1} \left(2 v_r^2-3 r^2 \Omega ^2\right)\Biggr] \\
            cr_{\vartheta} &= - G \Omega  r^{-2 \alpha -5} \Biggl[G \left\{(3 \alpha -1) k^2 r^6 M_{\mathrm{MBH}}^{\alpha - 1}+(3 \alpha +7) k M_{\mathrm{MBH}}^{\frac{\alpha +1}{2}} r^{\alpha +3}+8 M_{\mathrm{MBH}}^2 r^{2 \alpha }\right\} \\ & \quad +3 \alpha  k M_{\mathrm{MBH}}^{\frac{\alpha - 1}{2}} r^{\alpha +4} \left\{(\alpha +1) v_r^2-r^2 \Omega ^2\right\}-9 M_{\mathrm{MBH}} r^{2 \alpha +1} \left(r^2 \Omega ^2-4 v_r^2\right)\Biggr]
        \end{aligned}
    \end{equation}
    \item Pop ($\boldsymbol{p}$):
    \begin{equation}
        \label{eq: pop_BW}
        \begin{aligned}
            p_r &= - G r^{-3 \alpha -8} \Biggl[G^2 \biggl\{\left(4 \alpha^2-5 \alpha +1\right) k^3 r^9 M_{\mathrm{MBH}}^{\frac{3 (\alpha - 1)}{2}}+\left(7 \alpha ^2-4 \alpha +15\right) k^2 M_{\mathrm{MBH}}^{\alpha} r^{\alpha +6} +22 M_{\mathrm{MBH}}^{3} r^{3 \alpha } \\ & \quad +\left(3 \alpha ^2+\alpha +36\right) k M_{\mathrm{MBH}}^{\frac{\alpha + 3}{2}} r^{2 \alpha +3} \biggr\} - G \Omega ^2 r^{\alpha +3} \biggl\{\alpha  (7 \alpha +1) k^2 r^6  M_{\mathrm{MBH}}^{\alpha - 1} +66 M_{\mathrm{MBH}}^{2} r^{2 \alpha } +\left(6 \alpha ^2+7 \alpha +57\right)  \\ & \quad \times k M_{\mathrm{MBH}}^{\frac{\alpha + 1}{2}} r^{\alpha +3}\biggr\} + v_r^2 r^{\alpha +1} \biggl\{G \biggl(\alpha  \left(11 \alpha ^2-10 \alpha -1\right) k^2 r^6 M_{\mathrm{MBH}}^{\alpha - 1} + 204 M_{\mathrm{MBH}}^{2} r^{2 \alpha } +2 k M_{\mathrm{MBH}}^{\frac{\alpha + 1}{2}} r^{\alpha +3} \\ & \quad \times \left(3 \alpha ^3+5 \alpha ^2+7 \alpha +57\right) \biggr) -6 \Omega ^2 r^{\alpha +3} \left(\alpha  \left(\alpha ^2+3 \alpha +2\right) k r^3 M_{\mathrm{MBH}}^{\frac{\alpha - 1}{2}}+60 M_{\mathrm{MBH}} r^{\alpha }\right)\biggr\}  +3 \Omega ^4 r^{2 \alpha +6} \\ & \quad \times \left\{ \alpha  (\alpha +2) k r^3 M_{\mathrm{MBH}}^{\frac{\alpha - 1}{2}}+15 M_{\mathrm{MBH}} r^{\alpha }\right\}  + v_r^4 r^{2 \alpha +2} \left\{\alpha  \left(\alpha ^3+2 \alpha ^2-\alpha -2\right) k r^3 M_{\mathrm{MBH}}^{\frac{\alpha - 1}{2}}+120 M_{\mathrm{MBH}} r^{\alpha }\right\} \Biggr] \\
            p_{\vartheta} &= 2 G v_r \Omega  r^{-2 (\alpha +3)} \Biggl[ G \biggl\{\alpha  (8 \alpha +1) k^2 r^6 M_{\mathrm{MBH}}^{\alpha - 1}+\left(6 \alpha ^2+13 \alpha +57\right) k M_{\mathrm{MBH}}^{\frac{\alpha + 1}{2}} r^{\alpha +3} +75 M_{\mathrm{MBH}}^2 r^{2 \alpha }\biggr\} \\ & \quad +2 \alpha  (\alpha +2) k M_{\mathrm{MBH}}^{\frac{\alpha - 1}{2}} r^{\alpha +4} \left\{(\alpha +1) v_r^2-3 r^2 \Omega ^2\right\}  +30 M_{\mathrm{MBH}} r^{2 \alpha +1} \left(4 v_r^2-3 r^2 \Omega ^2\right)\Biggr]
        \end{aligned}
    \end{equation}
    \item Lock ($\boldsymbol{l}$):
    \begin{equation}
        \begin{aligned}
            \label{eq: lock_BW}
            l_r &= G v_r r^{-3 \alpha} \Biggl[ G^2 \Biggl\{ \left(34 \alpha ^3-39 \alpha ^2+6 \alpha -1\right) k^3 M_{\mathrm{MBH}}^{\frac{3 (\alpha - 1) }{2}} + 6 \left(8 \alpha ^3+\alpha ^2+8 \alpha +43\right) k^2 M_{\mathrm{MBH}}^{\alpha} r^{\alpha - 3} + 584 M_{\mathrm{MBH}}^{3} r^{3\alpha - 9 } \\ & \quad + 3 \left(5 \alpha ^3+12 \alpha ^2+23 \alpha +272\right) k M_{\mathrm{MBH}}^{\frac{\alpha + 3}{2}} r^{2 (\alpha - 3)} \Biggr\} - G \Omega ^2 r^{\alpha} \Biggl\{ \alpha  \left(48 \alpha ^2+55 \alpha +27\right) k^2 M_{\mathrm{MBH}}^{\alpha - 1} +1872 M_{\mathrm{MBH}}^{2} r^{2 \alpha - 6} \\ & \quad + 3 \left(10 \alpha ^3+37 \alpha ^2+59 \alpha +468\right) k M_{\mathrm{MBH}}^{\frac{\alpha + 1}{2}} r^{\alpha - 3} \Biggr\} + v_r^2 r^{\alpha - 2} \Biggl\{ G \Biggl( \alpha  \left(26 \alpha ^3-\alpha ^2-16 \alpha -9\right) k^2 M_{\mathrm{MBH}}^{\alpha - 1} +1908 M_{\mathrm{MBH}}^{2} \\ & \quad \times r^{2 \alpha - 6} +2 \left(5 \alpha ^4+21 \alpha ^3+25 \alpha ^2+81 \alpha +468\right) k M_{\mathrm{MBH}}^{\frac{\alpha + 1}{2}} r^{\alpha - 3} \Biggr) - 10 \Omega ^2 r^{\alpha - 3} \left(\alpha  \left(\alpha ^3+6 \alpha ^2+11 \alpha +6\right) k M_{\mathrm{MBH}}^{\frac{\alpha - 1}{2}} \right. \\ & \quad \left. +360 M_{\mathrm{MBH}} r^{\alpha - 3}\right) \Biggr\} +15 \Omega ^4 r^{2 \alpha} \left(\alpha  \left(\alpha ^2+5 \alpha +6\right) k M_{\mathrm{MBH}}^{\frac{\alpha - 1}{2}} +90 M_{\mathrm{MBH}} r^{\alpha - 3}\right) \\ & \quad + v_r^4 r^{2 (\alpha - 2)} \left(\alpha  \left(\alpha ^4+5 \alpha ^3+5 \alpha ^2-5 \alpha -6\right) k M_{\mathrm{MBH}}^{\frac{\alpha - 1}{2}}+720 M_{\mathrm{MBH}} r^{\alpha - 3}\right) \Biggr] \\
            l_{\vartheta} &= -G \Omega  r^{1 - 3 \alpha} \Biggl[ G^2 \Biggl\{ \left(20 \alpha ^2-3 \alpha +1\right) k^3 M_{\mathrm{MBH}}^{\frac{3 (\alpha - 1)}{2}}+\left(35 \alpha ^2+24 \alpha +129\right) k^2 M_{\mathrm{MBH}}^{\alpha} r^{\alpha - 3} +172 M_{\mathrm{MBH}}^3 r^{3 (\alpha - 3)} \\ & \quad + 3 \left(5 \alpha ^2+9 \alpha +100\right) k M_{\mathrm{MBH}}^{\frac{\alpha + 3}{2}} r^{2 (\alpha - 3)} \Biggr\} - G \Omega ^2 r^{\alpha} \Biggl\{ \alpha  (35 \alpha +27) k^2 M_{\mathrm{MBH}}^{\alpha - 1} +396 M_{\mathrm{MBH}}^2 r^{2 (\alpha - 3)} \\ & \quad +3 \left(10 \alpha ^2+19 \alpha +117\right) k M_{\mathrm{MBH}}^{\frac{\alpha +1}{2}} r^{\alpha - 3} \Biggr\} + v_r^2 r^{\alpha - 2} \Biggl\{ G \left(\alpha  \left(55 \alpha ^2+62 \alpha +27\right) k^2 M_{\mathrm{MBH}}^{\alpha - 1} +2124 M_{\mathrm{MBH}}^2 r^{2 (\alpha - 3)} \right. \\ & \quad \left. +6 \left(5 \alpha ^3+22 \alpha ^2+47 \alpha +234\right) k M_{\mathrm{MBH}}^{\frac{\alpha +1}{2}} r^{\alpha - 3}\right) -30 \Omega ^2 r^{\alpha - 3} \left(\alpha  \left(\alpha ^2+5 \alpha +6\right) k M_{\mathrm{MBH}}^{\frac{\alpha - 1}{2}}+90 M_{\mathrm{MBH}} r^{\alpha - 3 }\right) \Biggr\} \\
            & \quad +15 \Omega ^4 r^{2 \alpha} \left(\alpha  (\alpha +2) k M_{\mathrm{MBH}}^{\frac{\alpha - 1}{2}}+15 M_{\mathrm{MBH}} r^{\alpha - 3 }\right) \\
            & \quad +5 v_r^4 r^{2 (\alpha - 2)} \left(\alpha  \left(\alpha ^3+6 \alpha ^2+11 \alpha +6\right) k M_{\mathrm{MBH}}^{\frac{\alpha - 1}{2}}+360 M_{\mathrm{MBH}} r^{\alpha - 3 }\right) \Biggr]
        \end{aligned}
    \end{equation}
\end{itemize}

Most of these higher derivatives will be measurable only closer to the NSC's centre where the MBH's potential is dominant. Therefore, in those regions even though we will be able to measure the higher derivatives, it may not be possible to distinguish between different values of $\alpha$. However, as we go away from the center, higher derivatives start becoming weaker in magnitude, and will not be measurable. As a result, the precision of measuring $\alpha$ will diminish. Therefore, there exists a sweet spot where we should be able to constrain $\alpha$ well.

\subsection{Orbits in Plummer Potential}\label{subsec: app_PP}
For binary mergers in GCs, we use the Plummer profile \cite{Plummer1911, 1974A&A....37..183A} defined as:
\begin{equation}
    \label{eq: rho_PP}
    \rho (r) = \frac{3 M_0}{4 \pi a_{\mathrm{p}}^3}\left(1 + \frac{r^2}{a_{\mathrm{p}}^2} \right)^{-5/2}
\end{equation}
where $M_0$ is the total mass of the cluster and $a_{\mathrm{p}}$ is the Plummer radius. The enclosed mass, gravitational potential, and acceleration are given by:
\begin{align}
    \label{eq: M_enc_PP}
    M_{\mathrm{enc}} (r) &= M_0 \frac{r^3}{(r^2 + a_{\mathrm{p}}^2)^{3/2}} \\
    \label{eq: Phi_PP}
    \Phi (r) &= -\frac{G M_0}{\sqrt{r^2 + a_{\mathrm{p}}^2}} \\
    \label{eq: acc_PP}
    a_r &= - G M_0 \frac{r}{\left(a_{\mathrm{p}}^2+r^2\right)^{3/2}}
\end{align}

Equations \eqref{eq: om_BW} - \eqref{eq: dombydr_BW} retain their same form with the new definition of variables which are now in the context of Plummer Potential. We once again follow the recursive process of calculating $\boldsymbol{j, s}$ to get the radial and angular components of these quantities as:
\begin{itemize}
    \item Jerk ($\boldsymbol{j}$):
    \begin{equation}
        \label{eq: jerk_PP}
        \begin{aligned}
            j_r &= - G M_0 \frac{v_r \left(a_{\mathrm{p}}^2-2 r^2\right)}{\left(a_{\mathrm{p}}^2+r^2\right)^{5/2}} \\
            j_{\vartheta} &= - G M_0 \frac{r\Omega}{\left(a_{\mathrm{p}}^2+r^2\right)^{3/2}}
        \end{aligned}
    \end{equation}
    \item Snap ($\boldsymbol{s}$):
    \begin{equation}
        \label{eq: snap_PP}
        \begin{aligned}
            s_r &= G M_0 \frac{r \left[G M_0 \left(a_{\mathrm{p}}^2-2 r^2\right)+3 \sqrt{a_{\mathrm{p}}^2+r^2} \left\{v_r^2 \left(3 a_{\mathrm{p}}^2-2 r^2\right)+r^2 \Omega ^2 \left(a_{\mathrm{p}}^2+r^2\right)\right\}\right]}{\left(a_{\mathrm{p}}^2+r^2\right)^4} \\
            s_{\vartheta} &= 6 G M_0  \frac{r^2 v_r \Omega }{\left(a_{\mathrm{p}}^2+r^2\right)^{5/2}}
        \end{aligned}
    \end{equation}   
\end{itemize}
In this case, even LOSS is barely measurable in LISA as well as DECIGO for 4 years of observation time. Therefore, we do not include the expressions for higher derivatives: $\boldsymbol{cr, p}$ etc.

\section{Additional Figures}\label{sec: supp_fig}
We plot a series of figures to complement those presented in the paper. 

\begin{figure*}[ht!]
    \centering
    \begin{subfigure}[b]{0.45\textwidth}
        \centering
        \includegraphics[width = 0.95\linewidth]{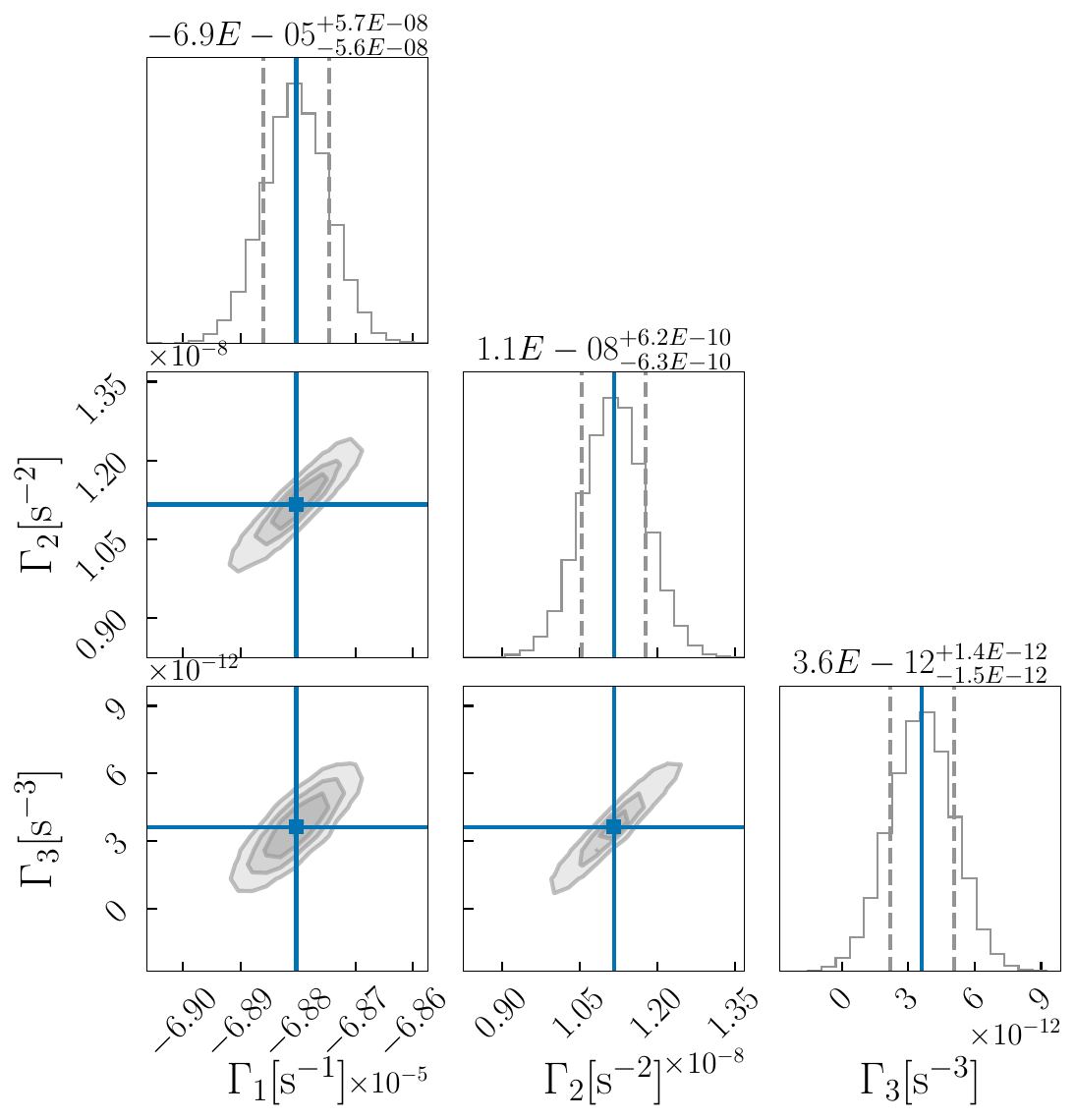}
    \end{subfigure}
    \begin{subfigure}[b]{0.45\textwidth}
        \centering
        \includegraphics[width = 0.95\linewidth]{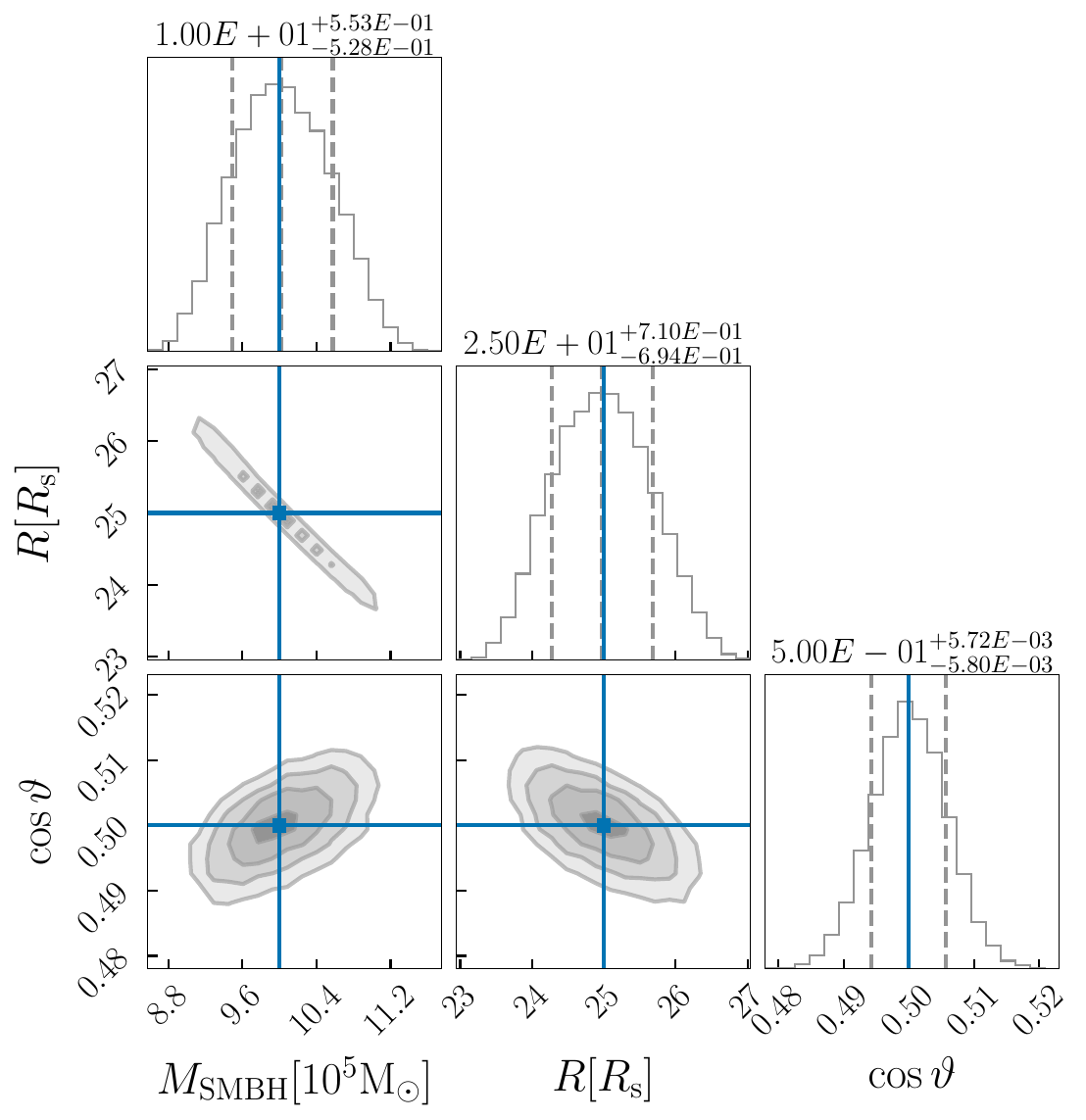}
    \end{subfigure}
    \caption{{\it Left Panel}: Corner plot showing the recovery of kinematic parameters: LOSA, LOSJ, and LOSS, for a $30-30 \, {\rm M}_{\odot}$ binary in a circular orbit of radius $25 \, R_{\rm s}$ around a $10^6\, {\rm M}_{\odot}$ SMBH, at $100\, {\rm Mpc}$, in ET. {\it Right Panel}: Corner plot showing the recovery of environmental parameters: $M_{\rm SMBH}$, $R$, and $\cos \vartheta$. The error bars provided on the top of the squares of the corner plots, here, and in other corner plots below, correspond to $68\%$ confidence intervals.}
    \label{fig: CO_cos_theta_0.5}
\end{figure*}

Figure~\ref{fig: CO_cos_theta_0.5} shows the corner plots of the recovery of kinematic ({\it left panel}) and environmental ({\it right panel}) parameters for a $30-30 \, {\rm M}_{\odot}$ binary in circular orbit around an SMBH, in ET, with $\cos \vartheta = 0.5$. 

Figure~\ref{fig: CO_cos_theta_0.5} shows the corner plot showing the inference of $\{M_{\mathrm{MBH}}, r, \alpha\}$, and correlation among them, for a $30-30 \, \mathrm{M}_{\odot}$ binary in an NSC with BW potential, in DECIGO. 

We argued in the paper and Appendix~\ref{sec: amp_cor} that amplitude corrections to the CBC waveform due to time derivatives of the CoM's LOSV can be neglected. To illustrate this, we plot the relative contributions to the amplitude from negative PN orders containing imprints of LOSA and its time derivatives. We also present the relative contributions to the GW phase from these negative PN orders. In particular, the {\it Left Panel} of Figure \ref{fig: pc_ac_BW} shows the comparison between relative phase corrections due to time derivatives of LOSV (see the equations in Appendix~\ref{sec: phase_cor}), and the $3.5$PN term of the \texttt{TaylorF2} phase due to point-particle approximation. The {\it Right Panel} shows the variation of the relative amplitude correction. Both phase and amplitude corrections are plotted for the frequency range $f_l = 0.018 \, {\rm Hz}$ to $f_h = 10 \, {\rm Hz}$, which corresponds to an observation time of 4 years in DECIGO. At low frequencies, up to $\sim 0.03 \, {\rm Hz}$, all of the derivatives considered here give rise to phase corrections larger than the $3.5$PN term. However, most of them become considerably smaller than this term for $f \gtrsim 0.03$ Hz, except for the $-4$ PN correction which exceeds the $3.5$ PN term even up to $\sim 0.2 \, {\rm Hz}$. From the {\it Right Panel} of Figure \ref{fig: pc_ac_BW}, it is clear that the leading order (relative) contributions (see the equations in Appendix~\ref{sec: amp_cor}) to the amplitude, due to the derivatives of LOSV, are very small ($\ll 1$). Moreover, matched-filter-based approaches to GW detection and inference are mostly sensitive to the GW phase and significantly less sensitive to the amplitude. Therefore, we do not expect these amplitude corrections to affect the results of our analysis.

\begin{figure*}
    \centering
    \includegraphics[width=0.75\linewidth]{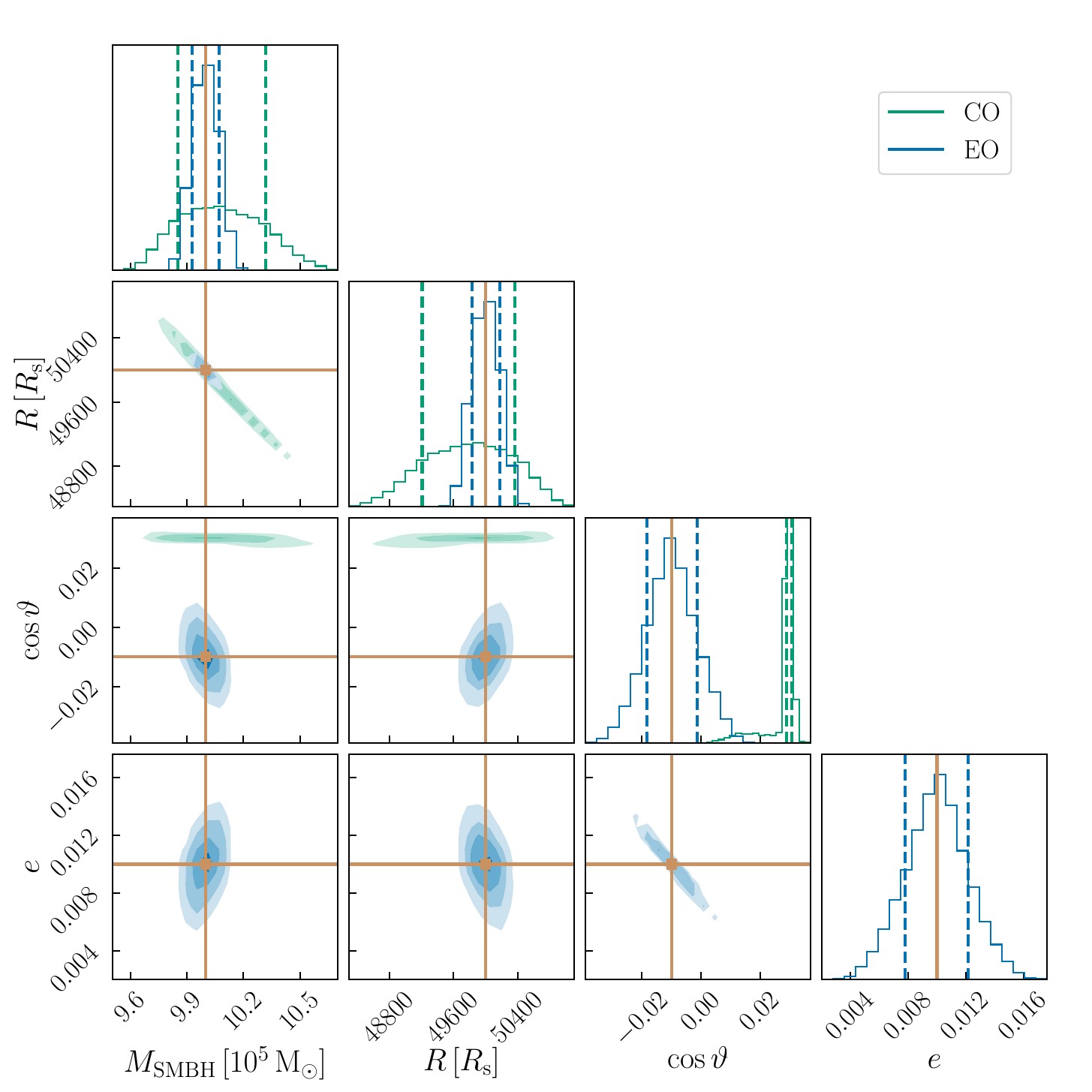}
    \caption{A comparison of the recoveries of environment parameters for Circular (Green) and Eccentric outer orbits (Blue). The CBC considered is a $100-100\, {\rm M_{\odot}}$ binary at $1 \, {\rm Gpc}$ in DECIGO in an eccentric outer orbit of semi-major axis $R = 5 \times 10^4 \,R_{\rm s}$ and eccentricity $e = 0.01$ around a $10^6 \, {\rm M_{\odot}}$ SMBH with $\cos \vartheta = - 0.01$ (i.e. $x = 1, r = R$). The orange lines represent the true values of the environment parameters. As can be seen, using an incorrect model gives a stealth bias.}
    \label{fig: stealth_bias}
\end{figure*}

Figure~\ref{fig: stealth_bias} shows a comparison of the recoveries of environment parameters for Circular (Pink) and Eccentric outer orbits (Orange) for $100-100 \, \rm M_{\odot}$ CBC considered in DECIGO.

\begin{figure}[ht!]
    \centering
    \includegraphics[width = 0.5\linewidth]{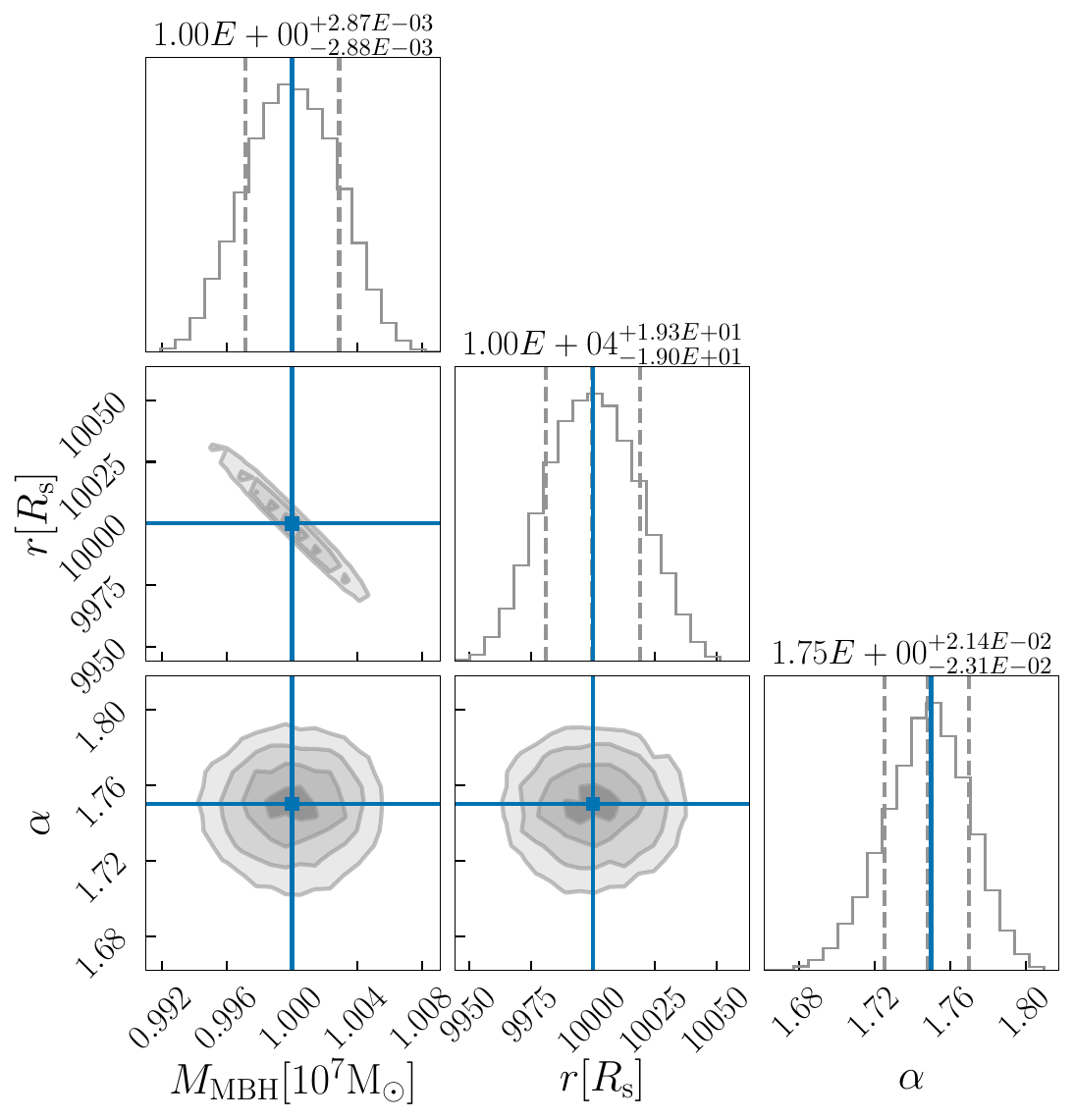}
    \caption{Corner plot showing the inference of $\{M_{\mathrm{MBH}}, r, \alpha\}$, and correlation among them, for a $30-30 \, \mathrm{M}_{\odot}$ binary in an NSC with BW potential, at $1\,\mathrm{Gpc}$, in DECIGO. The blue lines represent the true values.}
    \label{fig: EnvP_BW_DECIGO}
\end{figure}

\begin{figure*}[ht!]
    \centering
    \begin{subfigure}[b]{0.475\textwidth}
        \centering
        \includegraphics[width = 0.95\linewidth]{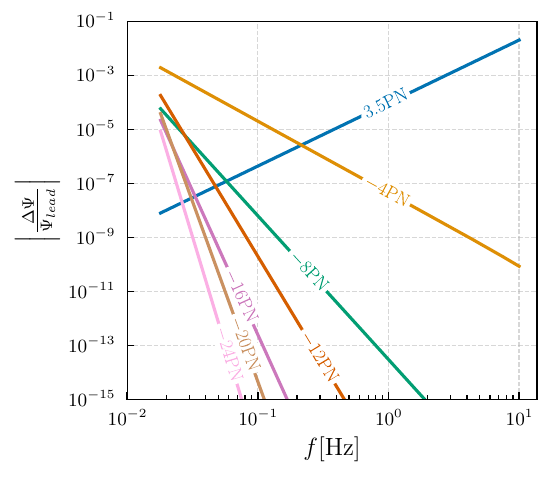}
    \end{subfigure}
    \begin{subfigure}[b]{0.475\textwidth}
        \centering
        \includegraphics[width = 0.95\linewidth]{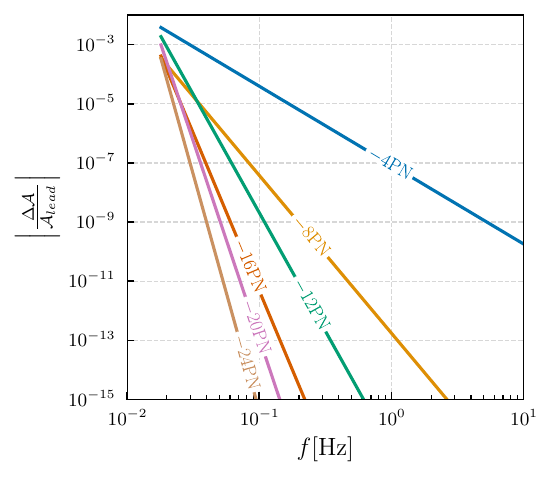}
    \end{subfigure}
    \caption{{\it Left Panel}: A comparison of the phase corrections due to time derivatives of the LOSV, with the $3.5$ PN term of the \texttt{TaylorF2} phase due to point-particle approximation, for the system considered in Figure \ref{fig: EnvP_BW_DECIGO}. {\it Right Panel}: The leading order amplitude corrections due to time derivatives of the LOSV. For the observation time of 4 years in DECIGO, $f_l = 0.018 \, \mathrm{Hz}$ and $f_h = 10 \, \mathrm{Hz}$.}
    \label{fig: pc_ac_BW}
\end{figure*}

\end{document}

%% file: preamble.tex
\usepackage[T1]{fontenc}
\usepackage[utf8]{inputenc}
\DeclareUnicodeCharacter{2009}{\nobreakspace}
\usepackage{lmodern}
\usepackage{times}
\usepackage[varg]{txfonts}
\usepackage[normalem]{ulem}
\usepackage{verbatim}
\usepackage[dvipsnames, usenames]{xcolor}
\definecolor{linkcolor}{rgb}{0.0,0.3,0.5}
\usepackage[hypertexnames=false, unicode, colorlinks=true, linkcolor=linkcolor,
citecolor=linkcolor, filecolor=linkcolor,urlcolor=linkcolor,
pdfusetitle]{hyperref}
\usepackage[all]{hypcap}
\usepackage{graphicx}
\usepackage{xspace}
\usepackage{amssymb}
\usepackage{bm} 
\usepackage{microtype}
\usepackage[english]{babel}